\documentclass[12pt]{emulateapj}
\usepackage{natbib}
\usepackage[normalem]{ulem} 
\newcommand{\etal}{et~al.}
\newcommand{\eg}{e.g.,}

\newcommand{\ie}{i.e.,}
\newcommand{\zsp}{\left< z_{\mbox{\small{sp}}} \right>}

\newcommand{\mum}{$\mu$m}

\newcommand{\lir}{$L_{\mbox{\scriptsize{IR}}}$}
\newcommand{\lirmath}{L_{\mbox{\scriptsize{IR}}}}
\newcommand{\lsun}{$L_\odot$}
\newcommand{\lsunmath}{L_\odot}
\newcommand{\msunmath}{M_\odot}
\newcommand{\msun}{$M_\odot$}

\newcommand{\zcl}{z_{\mbox{\scriptsize cl}}}
\newcommand{\rtwo}{r_{\mbox{\scriptsize 200}}}
\newcommand{\rvir}{r_{\mbox{\scriptsize virial}}}
\newcommand{\spitzer}{{\it Spitzer}}
\newcommand{\chandra}{{\it Chandra}}
\newcommand{\hst}{{\it HST}}
\newcommand{\herschel}{{\it Herschel}}
\newcommand{\Mtwo}{M_{\mbox{\scriptsize 200}}}

\newcommand{\cl}{IDCS~J1426.5+3508}

\def\spose#1{\hbox to 0pt{#1\hss}}
\def\simlt{\mathrel{\spose{\lower 3pt\hbox{$\mathchar"218$}}
     \raise 2.0pt\hbox{$\mathchar"13C$}}}
\def\simgt{\mathrel{\spose{\lower 3pt\hbox{$\mathchar"218$}}
     \raise 2.0pt\hbox{$\mathchar"13E$}}}

\shorttitle{The Era of Star Formation in Galaxy Clusters}
\shortauthors{Brodwin et al.}

\begin{document}

%% LaTeX will automatically break titles if they run longer than
%% one line. However, you may use \\ to force a line break if
%% you desire.

\title{The Era of Star Formation in Galaxy Clusters}
%\title{The Star Formation Rate in Galaxy Clusters at $1<z<1.5$ in the IRAC
%  Shallow Cluster Survey}

%% Use \author, \affil, and the \and command to format
%% author and affiliation information.
%% Note that \email has replaced the old \authoremail command
%% from AASTeX v4.0. You can use \email to mark an email address
%% anywhere in the paper, not just in the front matter.
%% As in the title, you can use \\ to force line breaks.

\author{M.~Brodwin\altaffilmark{1}, 
 S.~A.~Stanford\altaffilmark{2},
 Anthony~H.~Gonzalez\altaffilmark{3},
 G.~R.~Zeimann\altaffilmark{4},
 G.~F.~Snyder\altaffilmark{5},
 C.~L.~Mancone\altaffilmark{3},
 A.~Pope\altaffilmark{6},
 P.~R.~Eisenhardt\altaffilmark{7}, 
 D.~Stern\altaffilmark{7},
 S.~Alberts\altaffilmark{6},
 M.~L.~N.~Ashby\altaffilmark{5}, 
 M.~J.~I.~Brown\altaffilmark{8}, 
 R.-R.~Chary\altaffilmark{9},
 Arjun Dey\altaffilmark{10}, 
 A.~Galametz\altaffilmark{11},
 D.~P.~Gettings\altaffilmark{3},
 B.~T.~Jannuzi\altaffilmark{12}, 
 E.~D.~Miller\altaffilmark{13}
 J.~Moustakas\altaffilmark{14}
 \& L.~A.~Moustakas\altaffilmark{7}}

%% Notice that each of these authors has alternate affiliations, which
%% are identified by the \altaffilmark after each name.  Specify alternate
%% affiliation information with \altaffiltext, with one command per each
%% affiliation.

\altaffiltext{1}{Department of Physics and Astronomy, University of
  Missouri, 5110 Rockhill Road, Kansas City, MO 64110}
\altaffiltext{2}{University of California, Davis, CA 95616}
\altaffiltext{3}{Department of Astronomy, University of Florida, Gainesville, FL 32611}
\altaffiltext{4}{Department of Astronomy and Astrophysics, Pennsylvania State University, 525 Davey Laboratory, University Park, Pennsylvania 16802}
\altaffiltext{5}{Harvard--Smithsonian Center for Astrophysics, 60 Garden Street, Cambridge, MA 02138}
\altaffiltext{6}{Department of Astronomy, University of Massachusetts, Amherst, MA 01003}
\altaffiltext{7}{Jet Propulsion Laboratory, California Institute of Technology, Pasadena, CA 91109}
\altaffiltext{8}{School of Physics, Monash University, Clayton,
  Victoria 3800, Australia}
\altaffiltext{9}{Spitzer Science Center, MC 220--6, California Institute of Technology, 1200 East California Boulevard, Pasadena, CA 91125}
\altaffiltext{10}{National Optical Astronomy Observatory, 950 N.~Cherry Ave., Tucson, AZ 85719}
\altaffiltext{11}{INAF -- Osservatorio di Roma, Via Frascati 33, I-00040, Monteporzio, Italy}
\altaffiltext{12}{Steward Observatory, University of Arizona, 933 N.~Cherry Ave., Tucson, AX 85121}
\altaffiltext{13}{Kavli Institute for Astrophysics and Space Research, Massachusetts Institute of Technology, Cambridge, MA 02139}
\altaffiltext{14}{Department of Physics and Astronomy, Siena College, 515 Loudon Road, Loudonville, NY 12211}

%% Mark off your abstract in the ``abstract'' environment. In the
%% manuscript style, abstract will output a Received/Accepted line
%% after the title and affiliation information. No date will appear
%% since the author does not have this information. The dates will be
%% filled in by the editorial office after submission.

\begin{abstract}

  We analyze the star formation properties of 16 infrared-selected,
  spectroscopically confirmed galaxy clusters at $1 < z < 1.5$ from
  the \spitzer/IRAC Shallow Cluster Survey (ISCS).  We present new
  spectroscopic confirmation for six of these high-redshift clusters,
  five of which are at $z>1.35$.  Using infrared luminosities measured
  with deep \spitzer/MIPS observations at 24 \mum, along with robust
  optical+IRAC photometric redshifts and SED-fitted stellar masses, we
  present the dust-obscured star-forming fractions, star formation
  rates and specific star formation rates in these clusters as
  functions of redshift and projected clustercentric radius.  We find
  that $z\sim 1.4$ represents a transition redshift for the ISCS
  sample, with clear evidence of an unquenched era of cluster star
  formation at earlier times.  Beyond this redshift the fraction of
  star-forming cluster members increases monotonically toward the
  cluster centers. Indeed, the specific star formation rate in the
  {\it cores} of these distant clusters is consistent with field
  values at similar redshifts, indicating that at $z>1.4$
  environment-dependent quenching had not yet been established in ISCS
  clusters.  Combining these observations with complementary studies
  showing a rapid increase in the AGN fraction, a stochastic star
  formation history, and a major merging episode at the same epoch in
  this cluster sample, we suggest that the starburst activity is
  likely merger-driven and that the subsequent quenching is due to
  feedback from merger-fueled AGN.  The totality of the evidence
  suggests we are witnessing the final quenching period that brings an
  end to the era of star formation in galaxy clusters and initiates
  the era of passive evolution.

\end{abstract}

%% Keywords should appear after the \end{abstract} command. The uncommented
%% example has been keyed in ApJ style. See the instructions to authors
%% for the journal to which you are submitting your paper to determine
%% what keyword punctuation is appropriate.

%\keywords{clusters: globular, peanut---bosons: bozos}

 %% Others ?

%% From the front matter, we move on to the body of the paper.
%% In the first two sections, notice the use of the natbib \citep
%% and \citet commands to identify citations.  The citations are
%% tied to the reference list via symbolic KEYs. The KEY corresponds
%% to the KEY in the \bibitem in the reference list below. We have
%% chosen the first three characters of the first author's name plus
%% the last two numeral of the year of publication as our KEY for
%% each reference.

\section{Introduction}

Galaxies in the centers of nearby rich clusters and groups are
passive, with little or no ongoing star formation.  Most optical and
near-IR studies of the color and luminosity function evolution in
cluster galaxies are consistent with a model in which cluster galaxies
formed in short, vigorous bursts of star formation at high redshift
($z \ga 2$), and evolved passively thereafter (\eg\
\citealt{stanford98}; \citealt{blakeslee06}; \citealt{vandokkum07};
\citealt{eisenhardt08}, hereafter E08; \citealt{mei09};
\citealt{mancone10}).

In contrast, recent mid-infrared studies \citep[\eg][]{bai09,chung10}
have demonstrated that galaxies with high star formation rates (SFRs),
including Luminous and Ultraluminous Infrared Galaxies (LIRGs and
ULIRGs, defined as having $8-1000$ $\mu$m infrared luminosities, \lir,
of $10^{11} \lsunmath \le \lirmath < 10^{12} \lsunmath$, and $\lirmath
\ge 10^{12} \lsunmath$, respectively), typically reside in the
outskirts of present-day massive clusters.  This suggests both an
ongoing level of infall of gas-rich galaxies and groups, and a
mechanism to quench the prodigious star formation of such recently
accreted cluster members.  Strangulation \citep{larson80} --- the
stripping of galaxies' hot gas reservoirs via interaction with the
intracluster medium (ICM) --- is the long-timescale ($\sim$few Gyr)
mechanism typically invoked to explain the lack of subsequent star
formation as a recently accreted galaxy is starved of fuel.  At low
redshift ($z<0.1$) this environmental quenching is so effective that
the fraction of star-forming galaxies in clusters is still below that
in the field even at $3\, \rtwo$ \citep{chung11}, where $\rtwo$
($\approx \rvir$) is the radius of the cluster within which the
density is 200 times the critical density of the Universe.

Evolutionary studies have found a rapid growth in the frequency and
intensity of the SFR in clusters.  \citet{saintonge08} reported an
increase in the fraction of rapidly star-forming cluster galaxies up
to $z\sim 0.8$.  The cluster mass-normalized integrated SFR was found
to increase as roughly $(1+z)^{5}$ out to $z\sim 1$
\citep{bai07,bai09,krick09}, at least as rapidly as the field, albeit
from a lower base.  Several other authors have found corroborating
evidence of increased star formation activity in distant clusters out
to $z \sim 0.8$, including a rising incidence of LIRGs and ULIRGs
\citep[\eg][]{coia05, geach06, marcillac07, muzzin08, koyama08,
  haines09, tran09, smith10, webb13}.  However, in all of these $z<1$
studies cluster cores ($r \la 0.25 \,\rtwo$) still show evidence of
substantial quenching, with much lower central SFRs than are seen in
the field.

Studies in deep field surveys have also addressed the effect of local
density on star formation activity.  \citet{elbaz07} and
\citet{cooper08} found evidence that the relation between SFR and
local galaxy density reverses at $z\sim 1$, in the sense that the SFR
begins to increase with increasing density.  Recent studies have
reported LIRG-level IR luminosities in cluster galaxies at $z=1.46$
\citep{hilton10} and $z=1.62$ \citep{tran10}.  However, as
\citet{geach06} demonstrate, there is a significant variation amongst
clusters even at moderate redshifts ($z \sim 0.5$). 

The next step is to characterize the star formation properties of a
large, uniformly selected sample of galaxy clusters at redshifts well
beyond unity in order to study the epoch of cluster formation.  In
this paper, we study the SFR and specific star formation rates (sSFR)
in 16 infrared-selected, spectroscopically confirmed $1 < z < 1.5$
clusters from the \spitzer/IRAC Shallow Cluster Survey (ISCS; E08).
This large statistical sample --- consisting of 196 spectroscopic
cluster members at $z>1$ including 91 at $z>1.35$, supplemented by
robust photo-z members for complete sampling --- allows accurate mean
cluster properties to be determined, overcoming shot noise due to low
numbers of objects in individual cluster cores and systematic
variations in star formation history.  We use 24\mum\ \spitzer\ data
to directly probe the obscured star formation largely missed by
optical approaches, and reject contaminating AGN using both X-ray and
mid-IR methods.

In Section \ref{Sec: Data} we describe the ISCS cluster sample as well
as the extensive photometry, spanning X-ray to 24\mum\ wavelengths,
that is available for all of these clusters.  We describe the uniform
photometric redshifts used to identify these clusters, and present
spectroscopy for 6 newly confirmed $z>1$ clusters, 5 of which are at
$z>1.35$.  Calculations of stellar masses, total IR luminosities and
SFRs are described in Section \ref{Sec: Physical}.  Cluster membership
criteria and AGN rejection methods are described in Section \ref{Sec:
  Membership}.  In Section \ref{Sec: SFR}, the SFR and sSFR in the
ISCS clusters is presented, and the implications for cluster formation
are discussed in \textsection{\ref{Sec: Discussion}}.  We present our
conclusions in \textsection{\ref{Sec: Conclusions}}.  We use Vega
magnitudes, the \citet{chabrier03} initial mass function (IMF), and
the WMAP7 cosmology of $(\Omega_\Lambda, \Omega_M, h) = (0.728, 0.272,
0.704)$ of
\citet{komatsu11}.

\newpage
\section{Data}
\label{Sec: Data}
\subsection{IRAC Shallow Cluster Survey}
The IRAC Shallow Cluster Survey (ISCS; E08) is a wide-field
infrared-selected galaxy cluster survey carried out using the
\spitzer/IRAC Shallow Survey imaging \citep{eisenhardt04} of the 8.5
deg$^2$ Bo\"otes field of the NOAO Deep, Wide-Field Survey
\citep[NDWFS;][]{ndwfs99}.  The clusters are identified via a wavelet
search algorithm operating on photometric redshift probability
distribution functions for 4.5\mum-selected galaxies in thin redshift
slices over the redshift range $0<z<2$.  The ISCS sample contains 335
clusters and groups in an area of 7.25 deg$^2$, 106 of which are at
$z>1$.

Details of the photometric redshifts are given in \citet[][hereafter
B06]{brodwin06_ISS} and a full description of the cluster search and
spectroscopy for a dozen $z>1$ ISCS clusters is given in E08 (also see
\citealt{stanford05}; B06; \citealt{elston06}). In \citet{brodwin11}
we presented our most distant cluster at that time, ISCS~J1432.4+3250
at $z = 1.49$, though a new, deep extension to the survey has thus far
identified two more distant clusters, at $z=1.75$
\citep{stanford12,brodwin12,gonzalez12} and $z=1.89$
\citep{zeimann12}. 

Here we focus on sixteen ISCS clusters at $1 < z < 1.5$, listed in
Table \ref{Tab: sample}, that have deep multi-wavelength follow-up
observations from the X-ray to the mid-infrared and are
spectroscopically confirmed.  These clusters likely all have similar
halo masses, in the range $\sim (0.8-2) \times 10^{14}$~\msun.  This
statement is based on X-ray, weak-lensing and dynamical masses that
have been measured for a subset of them \citep{brodwin11, Jee11}, as
well as on a clustering analysis of the full ISCS sample
\citep{brodwin07}.  In a companion paper, \citet{alberts13} conduct a
\herschel/SPIRE stacking analysis using the full ISCS catalog.

\begin{deluxetable*}{lccccl}
  \tabletypesize{\normalsize} \tablecaption{High Redshift,
    Multiwavelength ISCS Cluster Sample \label{Tab: sample}}
  \tablewidth{0pt} \tablehead{ \colhead{} & \colhead{} &
    \colhead{} & \colhead{} &
    \colhead{Number of}  & \colhead{}\\\colhead{} & \colhead{R.A.} &
    \colhead{Decl.} & \colhead{Spectroscopic} &
    \colhead{Spectroscopic}  & \colhead{Additional}\\
    \colhead{ID} & \colhead{(J2000)} & \colhead{(J2000)} &
    \colhead{Redshift} &     
    \colhead{Members\tablenotemark{$\ddagger$}} & \colhead{References}
  } \startdata
ISCS J1429.2+3357                           &  14:29:15.16  & +33:57:08.5    &   1.059  &   8  &  1, 2 \\
ISCS J1432.4+3332                           &  14:32:29.18  & +33:32:36.0    &   1.113  &  26  &  1, 2, 3 \\
ISCS J1426.1+3403                           &  14:26:09.51  & +34:03:41.1    &   1.135  &  12  &  1, 2 \\
ISCS J1425.0+3520\tablenotemark{$\dagger$}  &  14:25:03.44  & +35:20:10.4    &   1.157  &   8  &  - \\ % no grism
ISCS J1426.5+3339                           &  14:26:30.42  & +33:39:33.2    &   1.164  &  14  &  1, 2 \\
ISCS J1434.5+3427                           &  14:34:30.44  & +34:27:12.3    &   1.238  &  19  &  1, 2, 4 \\
ISCS J1429.3+3437                           &  14:29:18.51  & +34:37:25.8    &   1.261  &  18  &  1, 2 \\
ISCS J1432.6+3436                           &  14:32:38.38  & +34:36:49.0    &   1.351  &  12  &  1, 2  \\
ISCS J1425.3+3428\tablenotemark{$\dagger$}  &  14:25:19.33  & +34:28:38.2    &   1.365  &  14  &  2 \\
ISCS J1433.8+3325\tablenotemark{$\dagger$}  &  14:33:51.14  & +33:25:51.1    &   1.369  &   6  &  2 \\
ISCS J1434.7+3519                           &  14:34:46.33  & +35:19:33.5    &   1.374  &  10  &  1, 2 \\
ISCS J1432.3+3253\tablenotemark{$\dagger$}  &  14:32:18.31  & +32:53:07.8    &   1.395  &  10  &  2 \\
ISCS J1425.3+3250\tablenotemark{$\dagger$}  &  14:25:18.50  & +32:50:40.5    &   1.400  &   6  &  2 \\
ISCS J1438.1+3414                           &  14:38:08.71  & +34:14:19.2    &   1.414  &  16  &  1, 2, 5, 6  \\
ISCS J1431.1+3459\tablenotemark{$\dagger$}  &  14:31:08.06  & +34:59:43.3    &   1.463  &   6  &  2 \\
ISCS J1432.4+3250                           &  14:32:24.16  & +32:50:03.7    &   1.487  &  11  &  2, 6  \\
\enddata
\tablenotetext{$\ddagger$}{See \textsection{\ref{Sec: Membership}} for
  the definition of cluster membership.}
\tablenotetext{$\dagger$}{New spectroscopic confirmation in this
  work.}  \tablerefs{$^1$E08; $^2$\citet{zeimann13};
  $^3$\citet{elston06}; $^4$B06; $^5$\citet{stanford05};
  $^6$\citet{brodwin11}.}
\end{deluxetable*}

\subsection{New $z>1$ ISCS Clusters}

Six new $z>1$ ISCS clusters spanning $1.157 < z < 1.464$ have been
spectroscopically confirmed using a combination of multi-object Keck
optical spectroscopy and slitless near-IR grism spectroscopy using the
Wide Field Camera 3 (WFC3) on \hst.  A few additional members were
confirmed in the AGES survey \citep{kochanek12}.  Table \ref{Tab:
  spectra} lists the coordinates, observation dates, exposure times
and redshifts for previously unpublished members identified via
ground-based optical spectroscopy.  The new members identified via
infrared \hst/WFC3 grism observations, also used in this analysis, are
presented in \citet{zeimann13}.

%\LongTables
\begin{deluxetable*}{lccllcc}
    \tabletypesize{\normalsize} \tablecaption{Spectroscopic Cluster Members\tablenotemark{1}\label{Tab: spectra}}
  \tablewidth{0pt} \tablehead{ \colhead{} & \colhead{R.A.} &
    \colhead{Decl.} & \colhead{} &
    \colhead{} & \colhead{} & \colhead{Exposure Time}\\
    \colhead{ID} & \colhead{(J2000)} & \colhead{(J2000)} &
    \colhead{Spec-z} &
    \colhead{Instrument} &
    \colhead{UT Date} & \colhead{(s)}
   } \startdata
\cutinhead{ISCS\_J1425.0+3520 $\left<z\right> = 1.157$}
J142503.5+352013  &  14:25:03.58  & +35:20:13.4  & 1.1570  &       Hectospec   &    AGES\tablenotemark{2}        &      AGES\tablenotemark{2}       \\ 
J142504.6+352114  &  14:25:04.62  & +35:21:14.1  & 1.1572  &       LRIS   & 2009 Apr 28 &  6 $\times$ 1800s           \\ 
J142505.7+352248  &  14:25:05.73  & +35:22:48.3  & 1.156   &       LRIS   & 2009 Apr 28 &  6 $\times$ 1800s           \\ 
J142507.4+351902  &  14:25:07.42  & +35:19:02.4  & 1.157   &       LRIS   & 2009 Apr 28 &  6 $\times$ 1800s           \\ 
J142507.6+352151  &  14:25:07.61  & +35:21:51.2  & 1.154   &       LRIS   & 2009 Apr 28 &  6 $\times$ 1800s           \\ 
J142510.7+352315  &  14:25:10.72  & +35:23:15.0  & 1.1588  &       LRIS   & 2009 Apr 28 &  6 $\times$ 1800s           \\ 
J142512.0+351839  &  14:25:12.00  & +35:18:39.1  & 1.157   &       LRIS   & 2009 Apr 28 &  6 $\times$ 1800s           \\ 
J142512.1+351955  &  14:25:12.18  & +35:19:55.2  & 1.1449  &       LRIS   & 2009 Apr 28 &  6 $\times$ 1800s           \\ 
\cutinhead{ISCS\_J1426.5+3339 $\left<z\right> = 1.164$}
J142619.7+333717  &  14:26:19.74  & +33:37:17.0  & 1.165   &       LRIS   & 2012 Apr 20 &  3 $\times$ 1140s           \\ 
J142631.2+334307  &  14:26:31.20  & +33:43:07.1  & 1.160   &       LRIS   & 2012 Apr 20 &  3 $\times$ 1140s           \\ 
J142633.4+334224  &  14:26:33.46  & +33:42:24.5  & 1.160   &       LRIS   & 2012 Apr 20 &  3 $\times$ 1140s           \\ 
\cutinhead{ISCS\_J1425.3+3428 $\left<z\right> = 1.365$}
J142511.3+342852  &  14:25:11.31  & +34:28:52.8  & 1.3759  &       LRIS   &   2006 Apr 5 &  7 $\times$ 1760s \\ 
J142516.0+343040  &  14:25:16.02  & +34:30:40.9  & 1.39    &       LRIS   &   2006 Apr 5 &  7 $\times$ 1760s \\ 
J142519.0+342807  &  14:25:19.05  & +34:28:07.2  & 1.3574  &       LRIS   &   2006 Apr 5 &  7 $\times$ 1760s \\ 
\cutinhead{ISCS\_J1433.8+3325 $\left<z\right> = 1.369$}
J143333.9+332602  &  14:33:33.98  & +33:26:02.9  & 1.377   &       DEIMOS   &   2007 Apr 19 &   4 $\times$ 1800s          \\
J143351.5+332645  &  14:33:51.55  & +33:26:45.9  & 1.3687  &       Hectospec &   AGES\tablenotemark{2}        &     AGES\tablenotemark{2}        \\
J143349.0+332603  &  14:33:49.05  & +33:26:03.3  & 1.39    &       LRIS   &      2006 Apr 4           &    7 $\times$ 1740s\\ 
\cutinhead{ISCS\_J1432.3+3253 $\left<z\right> = 1.395$}
J143211.5+325646  &  14:32:11.56  & +32:56:46.7  & 1.401   &       LRIS   &  2012 Apr 20          &    4 $\times$ 1740s          \\ 
J143216.5+325433  &  14:32:16.54  & +32:54:33.9  & 1.3921  &    Hectospec   &     AGES\tablenotemark{2}        &       AGES\tablenotemark{2}      \\
\cutinhead{ISCS\_J1425.3+3250 $\left<z\right> = 1.400$}
J142523.8+325001  &  14:25:23.85  & +32:50:01.7  & 1.41    &       LRIS   &  2012 Apr 20          &     4 $\times$ 1740s         \\ 
J142520.3+324701  &  14:25:20.34  & +32:47:01.7  & 1.3972  &       Hectospec  &      AGES\tablenotemark{2}       &      AGES\tablenotemark{2}       \\ 
\cutinhead{ISCS\_J1431.1+3459 $\left<z\right> = 1.463$}
J143110.8+350016  &  14:31:10.88  & +35:00:16.4  & 1.477   &       LRIS   &     2012 Apr 21       &  4 $\times$ 1740s            \\ 
\cutinhead{ISCS\_J1432.4+3250 $\left<z\right> = 1.487$}
J143225.1+325013  &  14:32:25.15  & +32:50:13.6  & 1.49    &       LRIS   & 2012 Apr 20   &    4 $\times$ 1740s   \\ 
J143225.1+325010  &  14:32:25.18  & +32:50:10.4  & 1.491   &       LRIS   & 2012 Apr 20   &    4 $\times$ 1740s   \\ 
\enddata
\tablenotetext{1}{{\small Only previously unpublished redshifts from
    ground-based telescopes are included.  See the Table 1 references
    for previously published members.  Additional new redshifts
    obtained with the \hst/WFC3 grism are presented in a companion
    paper \citep{zeimann13}}.}  \tablenotetext{2}{{\small See
    \citet{kochanek12} for a complete description of the AGES
    spectroscopy.}}
\end{deluxetable*} 

The Keck optical spectra were reduced using standard techniques,
including flat-fielding, fringe correction for LRIS red-side spectra,
cosmic ray rejection, wavelength calibration and stacking.  Spectral
features were identified in the one-dimensional spectra extracted in
IRAF\footnote{IRAF is distributed by the National Optical Astronomy
  Observatory, which is operated by the Association of Universities
  for Research in Astronomy (AURA) under cooperative agreement with
  the National Science Foundation.}, although all identified emission
lines were verified to be robust in the 2-D spectra. Redshifts for
star-forming galaxies were determined from a combination of
[OII]~$\lambda3727$ emission and the 4000~\AA\ break or overall
spectral shape, whereas redshifts for passive galaxies were secured
primarily via Ca HK absorption lines.

\begin{figure*}[bthp]
\plotone{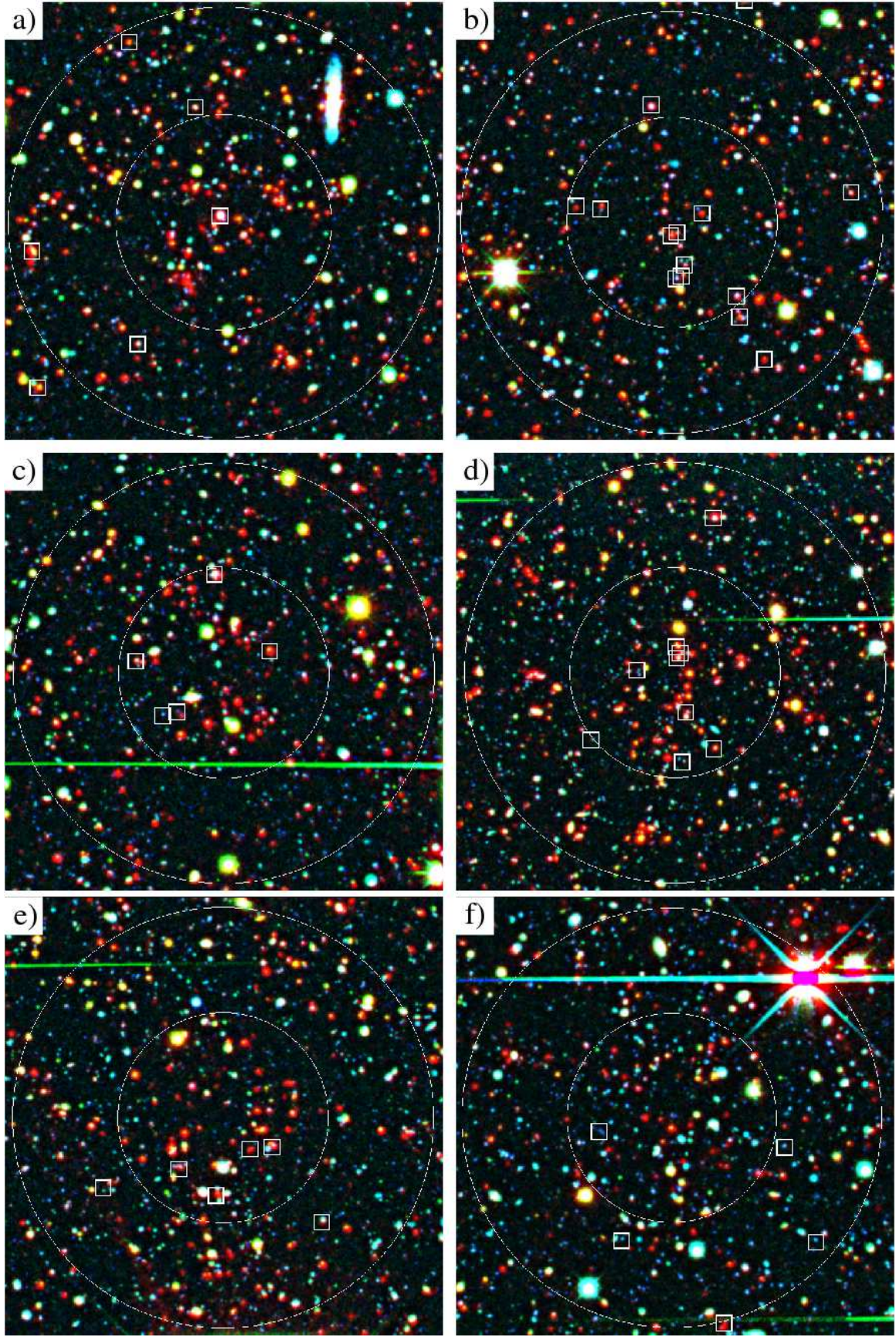}
\caption{4.1\arcmin $\times$ 4.1\arcmin optical $B_W$, $I$ and IRAC
  $4.5$ \mum\ pseudo-color images of 6 high-redshift clusters newly
  spectroscopically confirmed in this work.  The large circles denote
  radii of 0.5 and 1 Mpc and are centered on the wavelet detection
  centroids.  The white boxes indicate spectroscopic members; some
  spectroscopic members at radii between 1 and 2 Mpc are not shown.
  (a) ISCS~J1425.0+3520 at $\zsp = 1.157$; (b) ISCS~J1425.3+3428 at
  $\zsp = 1.365$; (c) ISCS~J1433.8+3325 at $\zsp = 1.369$; (d)
  ISCS~J1432.3+3253 at $\zsp = 1.395$; (e) ISCS~J1425.3+3250 at $\zsp
  = 1.400$; (f) ISCS~J1431.1+3459 at $\zsp = 1.463$.  The color image
  for the last cluster is made up of $B_W$, $R$ and IRAC $4.5$ \mum.}
\label{Fig: color image}
\end{figure*}

Optical/IRAC pseudo-color images of these new clusters, with the
spectroscopic members indicated, are shown in Figure \ref{Fig: color
  image}.  Although prominent red overdensities are present for most
of the clusters, the ISCS clusters are not red-sequence selected.  The
photometric redshift methodology includes bluer members, as is evident
in some of the panels of Figure \ref{Fig: color image}, and therefore
offers a selection that is less biased toward red-and-dead membership
than simple red-sequence surveys.  This is of crucial importance for
studies of the star formation activity in high-redshift clusters.

\subsection{Photometry and Photometric Redshifts}

\subsubsection{Optical and IRAC Data}

Deep, optical $B_WRI$ data from the NDWFS \citep{ndwfs99} is available
for all of these clusters (see B06 for more details).  In order to
match the larger PSFs of the \spitzer/IRAC photometry described below,
aperture--corrected 4\arcsec\ aperture fluxes were used.

The original 90s depth IRAC Shallow Survey was repeated three more
times as part of the \spitzer\ Deep, Wide-Field Survey
\citep[SDWFS;][]{ashby09} in \spitzer\ Cycle 4 (PID 40839), leading to
a factor of 2 increase in depth and a significantly more robust
catalog in terms of resistance to cosmic rays and instrumental
effects.  Combined with new, PSF-matched NDWFS optical catalogs (Brown
et al.~in prep), these data were used to compute new photometric
redshifts for the full 5 $\sigma$ 4.5\mum\ SDWFS sample consisting of
434,295 galaxies down to an aperture-corrected limit of [4.5] = 18.83
mag.

\newpage
\subsubsection{MIPS Data}

Imaging at 24 $\mu$m was obtained with the Multiband Imaging
Photometer for \spitzer\ (MIPS) in Cycle 3 (PID 30950) for all the
clusters in Table \ref{Tab: sample}.  The exposure time increased with
redshift from 12 to 48 min, corresponding to rms flux limits of 52
$\mu$Jy at $z=1$ to 12 $\mu$Jy at $z=1.5$, in order to uniformly
detect $3 \times 10^{11}$ \lsun\ LIRGs at S/N~$\ge 4$ for all clusters
in this sample.

Following generation of the Basic Calibrated Data (BCD) by the {\it
  Spitzer} Science Center pipeline, we flat-fielded our images using
scan mirror position-dependent flat fields derived from our science
data.  This is necessary because MIPS flat-fields are slightly
dependent on the position of scan mirrors\footnote{see
  http://irsa.ipac.caltech.edu/data/SPITZER/docs/mips/features/ for
  details.}. We also performed a jailbar correction on some science
data that presented a regular pattern of bars (due to the presence of
saturated pixels).  The images were then sky-subtracted and the final
mosaics were produced using MOPEX (with a drizzle scale of $0.7$ and a
pixel-resampling factor of $2$).

The MIPS data have an angular resolution of $5\farcs7$ FWHM, while the
relative astrometric accuracy derived by matching the 24 $\mu$m
sources with Two Micron All Sky Survey (2MASS) stars and the SDWFS
IRAC images is better than $0\farcs25$.  The sources discussed in this
paper are all unresolved at MIPS resolution; most are unresolved even
at IRAC resolution.  The MIPS source catalogs were generated by using
the positions of sources in the higher resolution IRAC images and
fitting groups of point sources using a singular value decomposition
technique at the positions of the IRAC sources to minimize the effect
of source confusion.  This is equivalent to a DAOPHOT-type approach
\citep{stetson87}, which is commonly adopted to obtain stellar
photometry in crowded
fields.  

\subsubsection{Chandra X-ray Data}

Several of these cluster positions had been previously imaged with
\chandra\ to 5-15 ks depths \citep{murray05}.  A Cycle 10 \chandra\
program added additional exposure time to bring the full sample to a
uniform exposure time of 40 ks.  Although the shallow X-ray exposures
were designed to study bright AGN, emission from the intracluster
medium (ICM) is detected for several of them.  A complete description
of the reduction of these data, along with ICM mass measurements for
two $z>1.4$ clusters in the present sample, is given in
\citet{brodwin11}.

\subsubsection{Matched Catalogs}

The various analyses in this work are based on cluster galaxy samples
selected in IRAC and/or MIPS bands from the global 4.5$\mu$m-limited
photometric redshift catalog described above.  Photometric redshift
and stellar mass fits use the optical and IRAC bands, with uniform
4\arcsec\ aperture fluxes measured at the positions of the SDWFS
4.5\mum\ sources.  Unlike in B06, the optical images were first
convolved to a uniform 1.35\arcsec\ PSF to produce more robust optical
colors, and to better match the native IRAC PSF.  All photometry was
corrected to total using a curve of growth analysis on bright,
unsaturated stars.  Since the SDWFS 4.5\mum\ catalog was used as a
positional prior to extract 24 $\mu$m fluxes in the MIPS images
\citep[\eg][]{magnelli09}, infrared luminosities (\lir) or limits are
measured for all sources.

\subsubsection{Photometric Redshifts}
\label{Sec: photoz}
The photometric redshift methodology adopted here is broadly similar
to that of B06, the main difference being that with the greater SDWFS
depth the 5.8 and 8.0\mum\ catalogs are included because they are
sensitive to non-local galaxy populations.  The extended \citet{cww}
and \citet{kinney96} templates used in B06 do not sample these
wavelengths, so in this work the models of \citet{polletta07} were
adopted.  Specifically, the templates employed include Ell5, Ell13,
S0, Sa, Sb, Sc, Sd, Spi4, and M82.  This subset of the
\citet{polletta07} templates, supplemented by the extended \citet{cww}
elliptical template, were empirically determined to provide an
excellent spectral basis for SDWFS galaxies at $0 < z < 2$
(i.e.~spanning the rest-frame wavelengths of $\sim$ 0.1-8 $\mu$m
probed by our filters).

The accuracy and precision of the new photometric redshifts are very
similar to those described in B06, with $\sigma/(1+z) \approx 0.06$
for 95\% of the galaxies.  The key improvement is that the photo-z
sample now extends to the SDWFS 5 $\sigma$ survey limit, corresponding
to 0.22 $L$* at $z=1.5$.  Unlike in B06, a neural-net approach was not
attempted for the bright AGNs as these are identified and removed
using our complementary data, as described below.  Figure \ref{Fig:
  photoz} shows the quality of the photometric redshifts for galaxies
on lines-of-sight toward the 16 clusters in Table \ref{Tab: sample}.
The slight bias to higher photometric redshifts evident at $z \la 0.5$
was not corrected as the present focus is on $z>1$ galaxies.  Larger
filled-in circles represent galaxies detected at S/N $\ge$ 4 in the
24\mum\ MIPS band, to which the rest of the discussion in this paper
will be limited.  Both the full and MIPS-detected line-of-sight galaxy
populations have redshift dispersions similar to the full field,
$\sigma/(1+z) = 0.064$ and $0.069$, respectively.  Spectroscopically
confirmed cluster members, denoted by stars, have a significantly
tighter dispersion, $\sigma/(1+z) = 0.039$.

\begin{figure}[bthp]
\epsscale{1.15} % turn on for emulateapj
\plotone{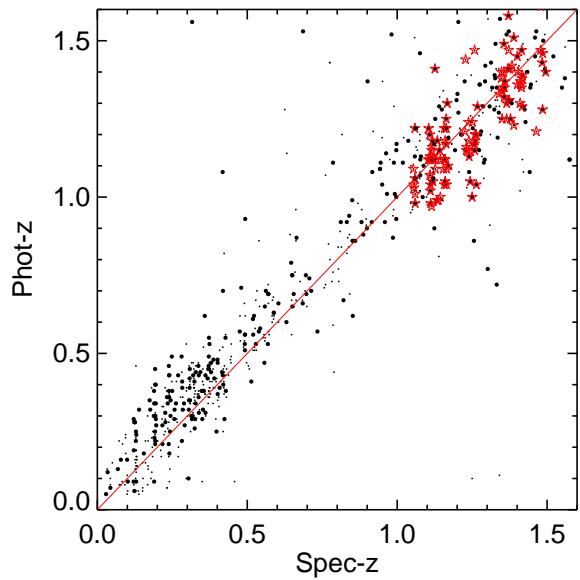}
\caption{Photometric redshift accuracy toward the 16 high--$z$
  clusters in Table 1.  AGN have been removed as described in
  \textsection{\ref{Sec: AGN}}. Larger symbols are MIPS-detected; the
  dispersion for this sample is the same as for the general galaxy
  population. The stars are $z>1$ spectroscopic cluster members.}
\label{Fig: photoz}
\end{figure}

As a further test of the reliability of the photometric redshifts,
Table \ref{Tab: PofZ} shows the fraction of galaxies for which the
spectroscopic redshift lies within the 1, 2 and 3$\sigma$ confidence
intervals, defined as the redshift regions that enclose the top
68.3\%, 95.4\%, and 99.7\% of the normalized area under the
photometric redshift probability distributions.  The photometric
redshift accuracy in these clusters fields is as good as or better
than that of the general SDWFS sample, particularly for the $z>1$
galaxies which form the basis of this work.

\begin{deluxetable}{lccc}
%\tabletypesize{\small}
\tablecaption{Confidence Level Statistics\tablenotemark{a}\label{Tab: PofZ}}
\tablewidth{0pt}
\tablehead{
  \colhead{Sample}& \colhead{$<1\sigma$}& \colhead{$<2\sigma$}&
  \colhead{$<3\sigma$}}\\
\startdata
Gaussian Expectation     &  68.3\% &  95.4\% & 99.7\%\\
SDWFS Main Galaxy Sample          &  73.8\% &  96.5\% & 99.3\%\\
This Work      &  74.6\% &  93.6\% & 98.6\%\\
This Work $z>1$&  81.1\% &  93.0\% & 98.7\%\\
\enddata
\tablenotetext{a}{Percentage of galaxies for which the spectroscopic redshift
  falls within the 1, 2, and 3$\sigma$ photometric redshifts
  confidence intervals.}
\end{deluxetable}

\section{Stellar Masses, Total Luminosities and Star Formation Rates}
\label{Sec: Physical}

\begin{figure}[bthp]
\epsscale{1.15} % turn on for emulateapj
\plotone{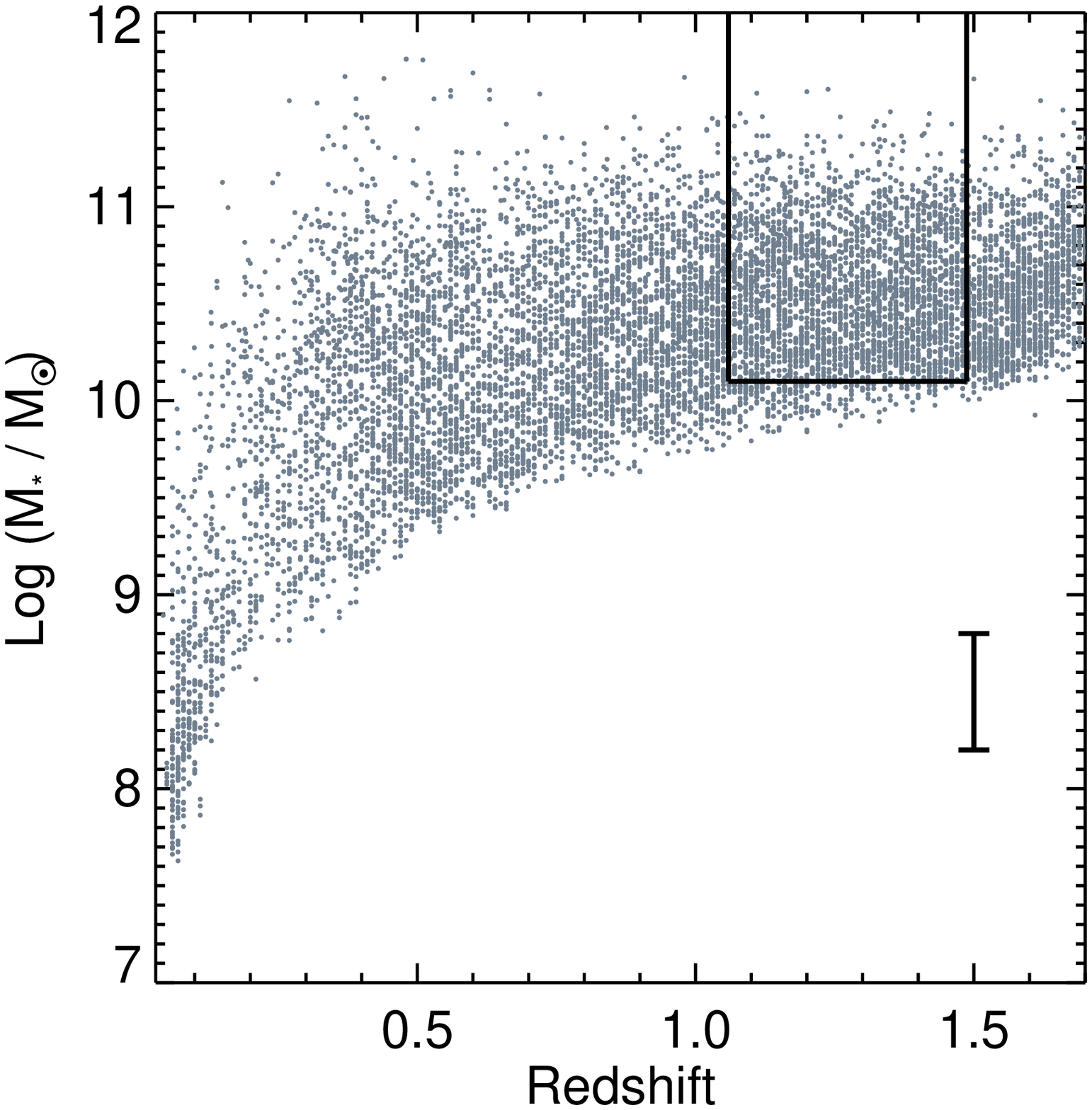}
\plotone{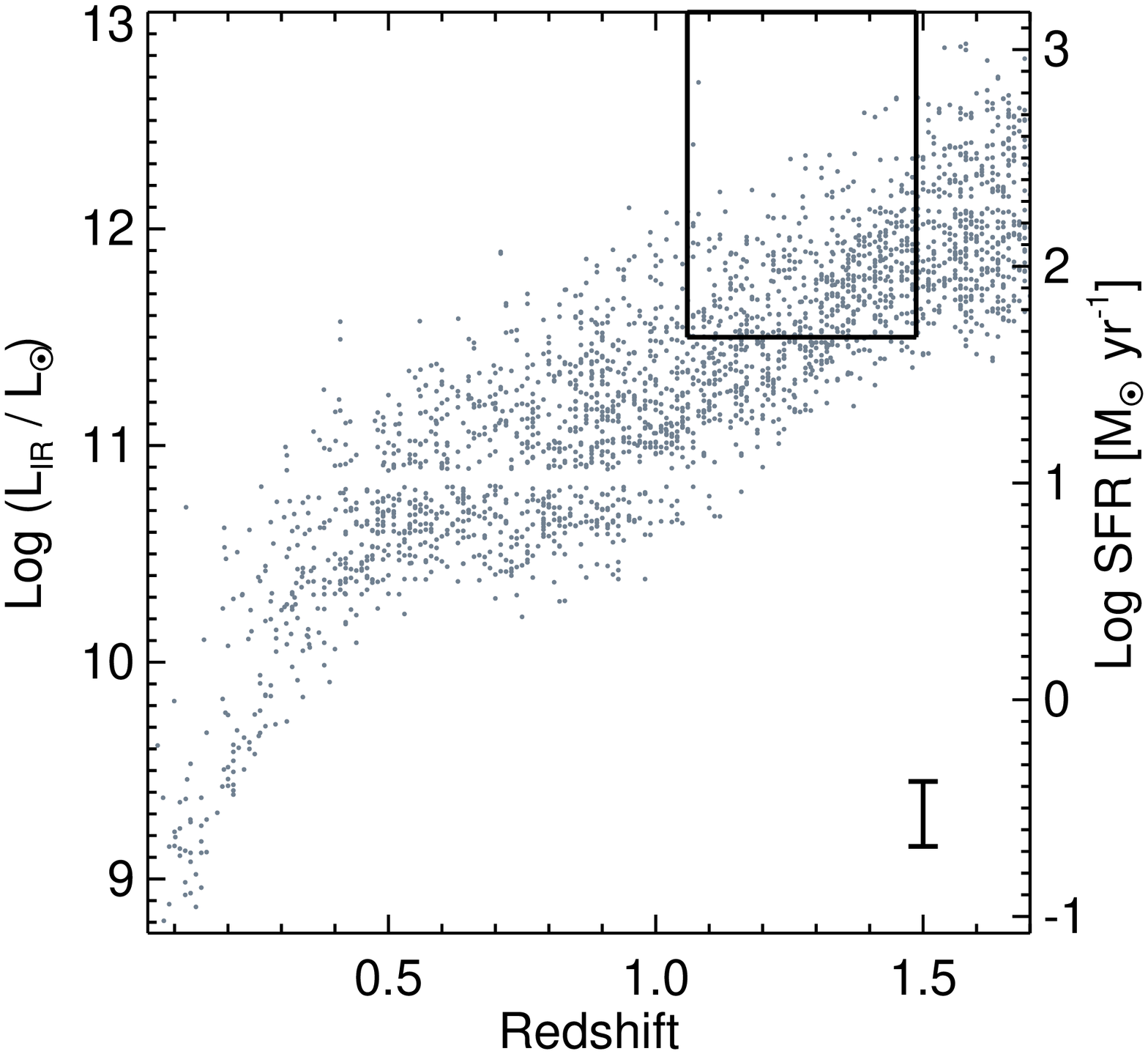}
\caption{{\it Upper panel:} Stellar masses along $5\arcmin \times
  5\arcmin$ lines-of-sight to the high-redshift clusters.  The
  selection region is enclosed by the box, corresponding to a stellar
  mass limit of $\log (M_\star / \msunmath) \ge 10.1$ for a Chabrier
  IMF.  {\it Lower panel:} Infrared luminosities in the same fields,
  with sampling complete to $\log (\lirmath / \lsunmath) \ge 11.5$.
  The redshifts in both panels are predominantly photometric.  The
  error bars are described in the text.}
\label{Fig: mass_lir}
\end{figure}

We estimate stellar masses using iSEDfit \citep{moustakas13}, a
Bayesian SED-fitting code that uses population synthesis models to
infer the physical properties of a galaxy given its observed broadband
SED.  We adopt the \citet{bc03} population synthesis models based on
the Padova isochrones, the {\sc stelib} \citep{leborgne03} stellar
library, and the \citet{chabrier03} IMF ranging from 0.1--100 \msun.
The upper panel of Figure \ref{Fig: mass_lir} shows the stellar masses
of all galaxies along the line-of-sight to these clusters.  For
uniformity across the $1<z<1.5$ cluster sample, we restrict the
stellar masses to $\log (M_\star / \msunmath) \ge 10.1$.  This limit
corresponds to the $\sim 80\%$ completeness level, though our cluster
member completeness is far higher over this redshift range given the
high masses of cluster galaxies and the flat luminosity function
\citep{mancone12}.  Although the individual iSEDfit mass errors are
typically $\le$ 0.2 dex, we adopt an error floor of 0.3 dex (indicated
by the error bar in the figure) to account for the systematic
uncertainty inherent in mass-fitting.

The total infrared luminosities of these galaxies are inferred from
the 24\mum\ fluxes using the \citet{chary01} templates.  While this
tends to overestimate \lir\ at high redshift ($z > 1.5$), particularly
for AGN-dominated ULIRGs \citep{murphy09,nordon10,rodighiero10}, it
provides an accurate (scatter $\sim 40\%$) estimate of \lir\ out to $z
= 1.5$ \citep{marcillac06,murphy09,Elbaz10}.  The MIPS data allow a
complete sample of total infrared luminosities down to \lir $ =
10^{11.5} $ \lsun\ for all our clusters (lower panel of Figure
\ref{Fig: mass_lir}), and we adopt this as the selection limit.  The
horizontal gap visible as a lack of sources with \lir $ \approx
10^{10.85} $ \lsun\ is an artifact stemming from the discreteness of
the \citet{chary01} templates.  Given our selection region, it does
not affect the present analysis.  The scatter is indicated by the
error bar in the figure.  We convert the total infrared luminosities
to star formation rates using the relation given in \citet{murphy11}.
This is defined assuming a \citet{kroupa01} IMF, and therefore has a
similar normalization to the Chabrier IMF used to calculate our
stellar masses.

The primary goals of this paper are to compare the SFR and sSFR in the
cluster outskirts and cores, and between clusters and the field.  In
all cases these SFRs have been derived using the same templates, so
these comparisons should be robust to small absolute deviations in SFR
compared with measurements at longer wavelengths (i.e.~with \herschel).

\section{Cluster Membership}
\label{Sec: Membership}
\subsection{Identification of Members}

Following E08, galaxies with robust spectroscopic redshifts are
identified as likely cluster members if they lie within a radius of 2
Mpc and their relative velocities are within $2000$ km s$^{-1}$ of the
systemic cluster velocity.  Galaxies without spectroscopy are
identified as cluster members via a constraint on the integral of
their normalized photometric redshift probability distribution
functions,
\begin{equation}
\int^{\zcl + 0.06(1+\zcl)}_{\zcl - 0.06(1+\zcl)} P(z)\, dz\ge 0.3.
\label{Eq: int_pz}
\end{equation}
$\zcl$ is the best-fit photometric redshift of the cluster, determined
by iteratively summing up the $P(z)$ functions for member galaxies
within 1 Mpc, re-identifying members and repeating the process to
convergence.  The positional centers are taken from the wavelet
algorithm used to initially identify the clusters, although we have
verified that using the Brightest Cluster Galaxy (BCG) position yields
similar results.

\subsection{Rejection of AGN}

\label{Sec: AGN} AGNs are problematic to include in this analysis,
both due to the difficulty of obtaining good photometric redshifts for
them, and because they bias the star formation rates inferred from the
infrared luminosity.  We therefore choose to omit them with the
understanding that the resulting star formation rates are formally
lower limits.  X-ray emitting AGN with 2--10 keV luminosities brighter
than $L_X > 10^{43}$ erg s$^{-1}$ were identified via a positional
match to the matched catalog, within a match radius that is the larger
of the IRAC PSF ($\approx 1.7\arcsec$) and the \chandra\ positional
uncertainty (which varies with off-axis angle).  We also identify AGN
via their power-law emission in the IRAC bands.  Objects with S/N $\ge
5$ in all four IRAC bands that fall in the \citet{stern05} AGN wedge
were deemed AGN.  Objects satisfying either of these AGN criteria are
removed from this analysis.

We have verified that the rejected AGN represent a relatively small
fraction of our cluster membership and have no strong redshift
dependence, as this could bias our primary result.  Only $\sim 4\%$ of
members satisfying our stellar mass cut are rejected as AGN within a
radius of 1 Mpc, with no apparent redshift trend (2.8\%, 6.0\% and
4.1\% in the 3 redshift bins used in \textsection{\ref{Sec: SFR}}).
Similarly, of the subset of members that are detected at 24$\mu$m,
only $\sim 12\%$ are rejected as AGN (11.1\%, 15.3\% and 9.9\% in
these redshift bins).  In the cluster cores, within projected radii of
0.5 Mpc, the rejected fractions are slightly higher ($\sim 7\%$ and
$\sim 18\%$ for all and 24$\mu$m-detected members, respectively), but
there is still no trend with redshift.  As we reject more AGN in the
cores we may possibly be underestimating the SFR in the cores relative
to the outskirts at all redshifts.  Since our results actually go the
other way, with higher activity in the cores, we conclude that the
rejection of AGN does not significantly bias our results with respect
to SFR trends in redshift or radius.

\section{Star Formation in High Redshift Clusters}
\label{Sec: SFR}

\subsection{24\mum-Detected Cluster Members}

\begin{figure*}[bthp]
\epsscale{1.15} % turn on for emulateapj
\plottwo{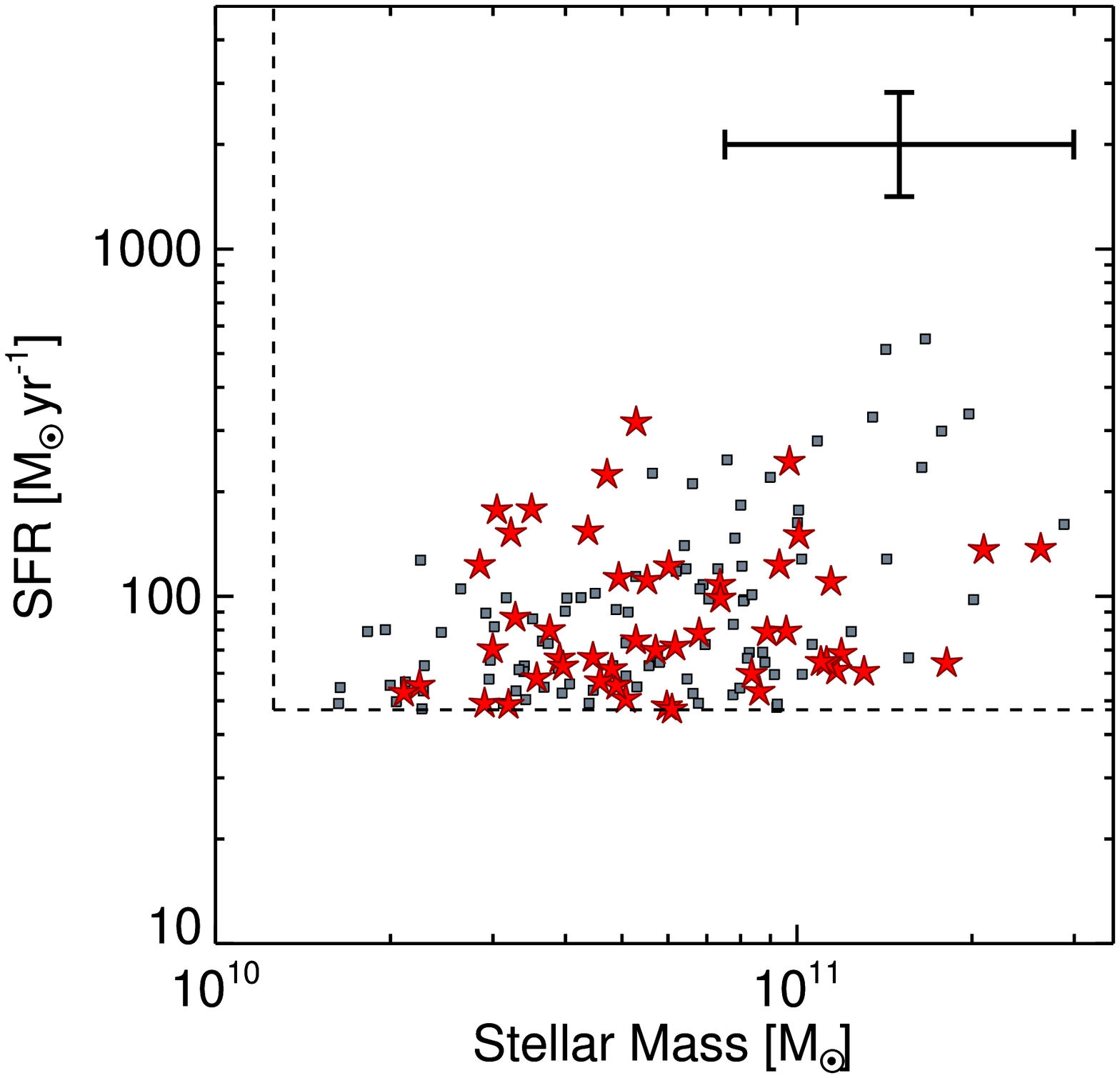}{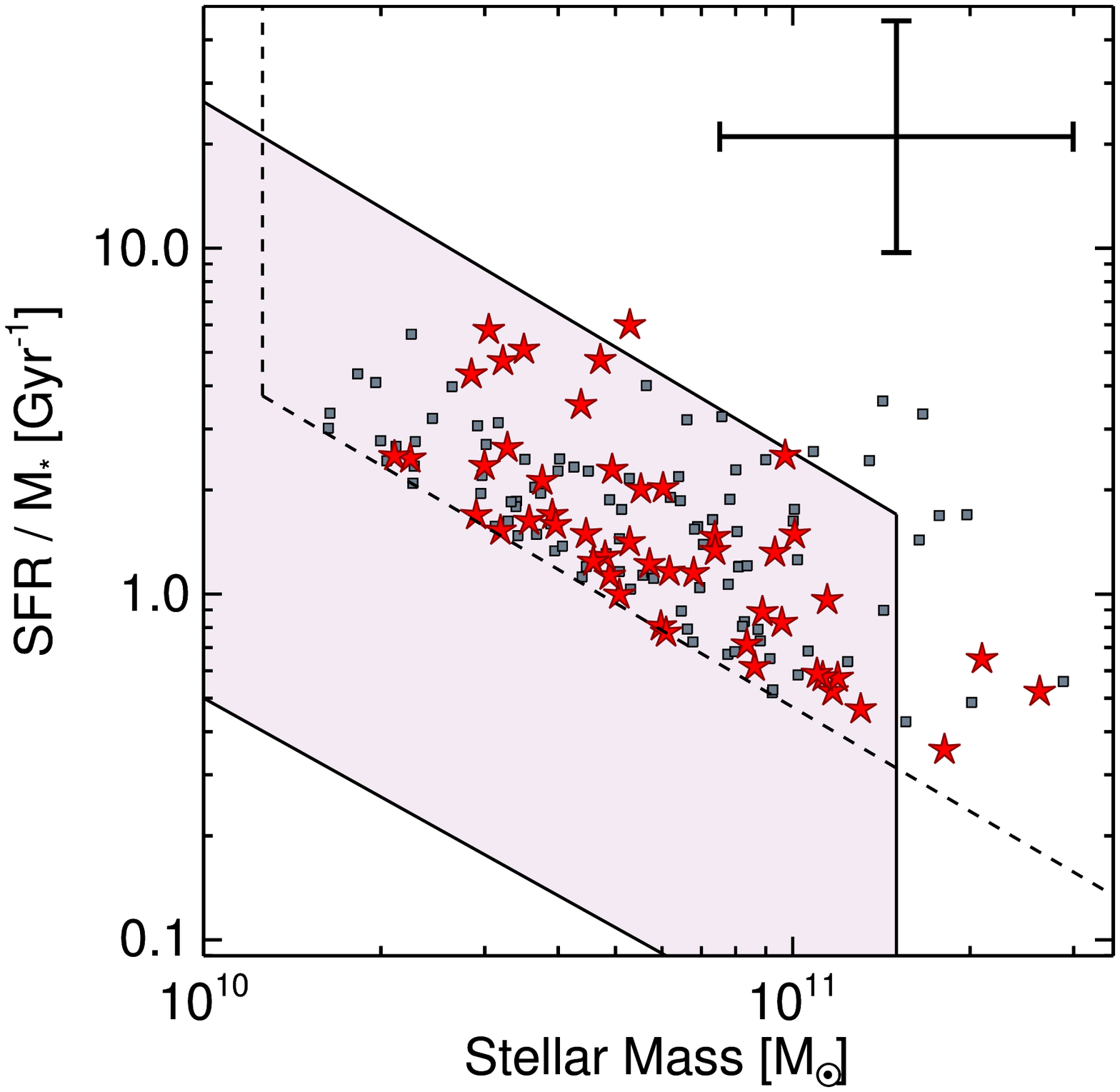}
\caption{Star formation ({\it left panel}) and specific star formation
  rates ({\it right panel}) vs.~stellar mass for the high-redshift
  cluster sample for galaxies detected at 24\mum\ with S/N $\ge 4$.
  The squares and stars represent photometric and spectroscopic
  redshift members, respectively.  The large error bars indicate the
  systematic uncertainties.  For comparison, the shaded region
  indicates the sSFR properties of the field galaxy sample of
  \citet{santini09}.  The dashed lines delineate the selection limits
  in stellar mass, SFR and sSFR.}
\label{Fig: ssfr_vs_mass}
\end{figure*}

Figure \ref{Fig: ssfr_vs_mass} shows the SFR ({\it left panel}) and
sSFR ({\it right panel}) as a function of stellar mass for galaxies
above the stellar mass and \lir\ limits given in
\textsection{\ref{Sec: Physical}}. The small boxes are galaxies
satisfying the cluster membership criteria defined in
\textsection{\ref{Sec: Membership}}, and the stars indicate galaxies
for which cluster membership has been spectroscopically confirmed.
The large error bars indicate the systematic errors.  For the stellar
masses these are conservatively estimated to be 0.3 dex, accounting
for IMF variations \citep[\eg][]{bell03}.  For the SFRs the systematic
error is taken to be 40\%, based on a comparison with \herschel\
far-IR measurements \citep{Elbaz10}.

The region of the sSFR plot populated by the field galaxy sample of
\citet{santini09} at $1<z<1.5$, adjusted to our choice of IMF, is
shown for comparison.  The MIPS data for this small field survey is
much deeper and hence probes to lower SFR.  However, to our SFR
sensitivity limit the cluster members, both spectroscopic and
photometric, have specific star formation rates very similar to these
field galaxies.  The spectroscopically confirmed cluster members span
a similar range in stellar mass, SFR and sSFR as the photometric
redshift members.  Given the small photometric redshift error for
cluster members (\textsection{\ref{Sec: photoz}}), the integrated
cluster SFR measurements, using spectroscopic redshifts where
available and photometric redshifts otherwise, should be
robust. 

\subsection{Star Formation vs.~Stellar Mass and Redshift}

\begin{figure}[bthp]
\epsscale{1.15} % turn on for emulateapj
\plotone{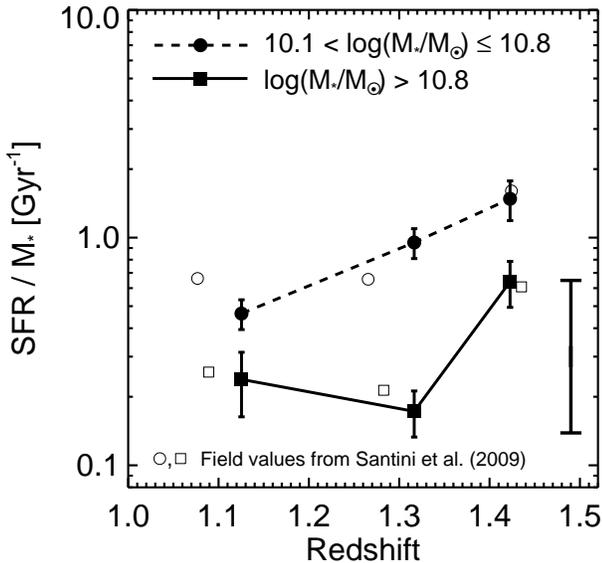}
\caption{Mean sSFR of central cluster galaxies, within projected radii
  of 500 kpc, in bins of stellar mass and redshift.  The errors on the
  points are from bootstrap resampling the galaxies in each bin, and
  the large error bar at right indicates the systematic uncertainty in
  the sSFR. For comparison, the open symbols show the sSFR the field
  galaxy sample of \citet{santini09} in the same mass and redshift
  bins.}
\label{Fig: mean_ssfr}
\end{figure}

The mean sSFR of the central cluster galaxies, within projected radii
of 500 kpc, are plotted in Figure \ref{Fig: mean_ssfr} in bins of
stellar mass and redshift.  The mean sSFR is defined here as the sum
of the SFRs divided by the sum of the stellar masses in the mass bin.
Objects undetected above 4 $\sigma$ at 24 \mum\ in any bin are
assigned the median 24 \mum\ flux of all such formally-undetected
objects in that bin.  This catalog-space median stacking is
complementary to the more common image-space stacking employed in a
companion paper \citep{alberts13}.
We have verified that none of the main results in this work depend on
the flux we infer for the undetected sources --- our primary
conclusions are unchanged if the fluxes of all galaxies with S/N $ <
4~\sigma$ MIPS detections are set to zero.  For comparison we plot the
sSFR of field galaxies from \citet[open symbols]{santini09} in the
same mass bins.  In this field study, SFRs for galaxies undetected at
24\mum\ are calculated from opt/NIR SED fits.

Although the methodologies differ in detail, the cluster and field
samples share some key characteristics.  The cluster galaxies appear
to lie in the same region of the sSFR--stellar mass plane as do the
star-forming field galaxies.  Similarly, the same evolutionary trend
is apparent in both cluster and field samples, with the sSFR
increasing over $1<z<1.5$.  The increase appears to be particularly
rapid for the more massive cluster galaxies above $z\ga 1.3$, perhaps
indicating that vigorous star formation is occurring in the massive
central galaxies in these clusters at levels comparable to the field.

\subsection{Star Formation vs.~Radius and Redshift}

\begin{figure}[bthp]
\epsscale{1.15} % turn on for emulateapj
\plotone{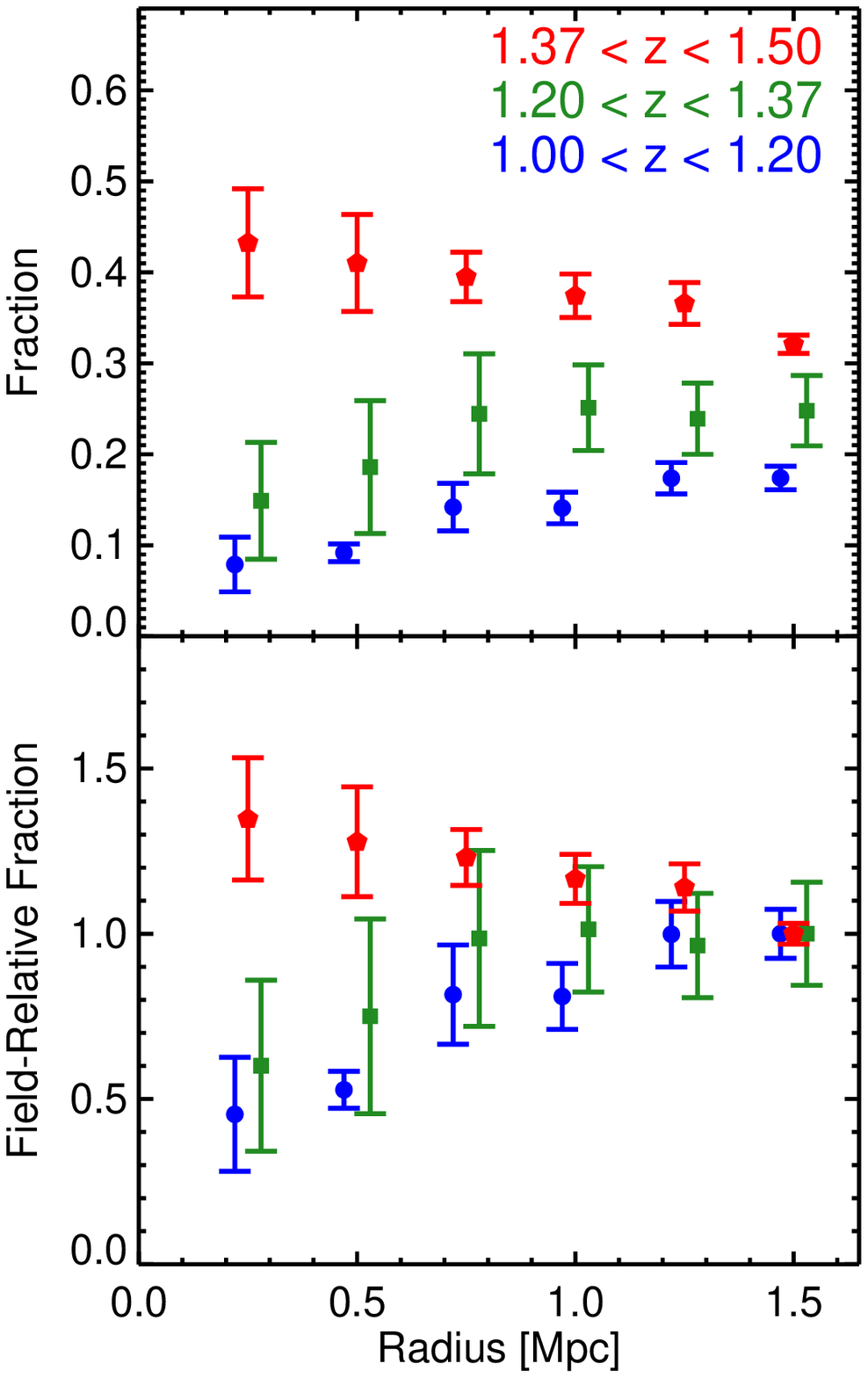}
\caption{{\it Upper panel:} Relative fraction of star-forming cluster
  members with \lir~$ \ge 10^{11.5}$ \lsun\ vs.~projected
  clustercentric radius for members with $\log (M_\star / \msunmath)
  \ge 10.1$.  {\it Lower panel:} Field-relative fractions, where the
  field fractions are taken to be the values at a radius of 1.5 Mpc.}
\label{Fig: fraction}
\end{figure}

To probe the effect of environment on star formation in $z>1$
clusters, we plot in Figure \ref{Fig: fraction} ({\it upper panel})
the fraction of cluster members with \lir~$ \ge 10^{11.5}$ \lsun\
vs.~projected clustercentric radius.  The sample is divided into three
redshift bins, chosen to have roughly equal numbers of clusters in
each bin, and the errors are estimated via bootstrap resampling.

In the lowest redshift bin ($1.0 < z < 1.2$) the LIRG fraction within
$\rtwo$ is roughly consistent with that seen at $z\sim 0.8$ by
\citet{finn10} in the ESO Distant Cluster Survey
\citep[EDisCS;][]{white05}, although that work probes to lower IR
luminosities ($\sim 10^{11}$ \lsun).  At the other extreme, the most
distant ISCS clusters have a LIRG fraction similar to, or even
somewhat higher than, IR-selected cluster ClG~J0218.3-0510 at $z=1.62$
\citep{tran10}.  However, as demonstrated below, there is significant
cluster to cluster variation in star formation properties, so
comparisons of individual clusters should be interpreted with caution.

For the ISCS clusters in the two bins at $z<1.37$ the fraction of
star-forming members drops significantly from a radius of about 1.5
Mpc, which is outside the virial radius ($\rtwo \sim 1$ Mpc) for this
sample, to the inner 250 kpc.  This is expected due to the quenching
of star formation in the central regions of clusters.  Indeed,
\citet{muzzin12} find a very similar star formation-radius trend in
the GCLASS sample at $z\sim 1$.

However, in the highest redshift bin ($1.37 < z < 1.50$) we find the
star-forming fraction does not drop, but rather {\em rises} with
decreasing radius right into the cluster cores.  This is consistent,
in an evolutionary sense, with the observation of \citet{muzzin12}
that the fraction of post-starburst galaxies in lower-redshift ($z\sim
1$) clusters increases toward the core.  Indeed, there is sufficient
time between $z\sim 1.5$ and $z\sim 1.0$ ($\sim 1.5$ Gyr) for a
significant fraction of the ISCS starbursts to evolve into
post-starbursts, but only if their star formation is rapidly quenched.

The upper panel of Figure \ref{Fig: fraction} also shows that the
fraction of star-forming galaxies in the field, taken here to be the
values observed at a radius of 1.5 Mpc, is also increasing with
redshift.  To better isolate the cluster-specific evolution, the
fractions in each redshift bin are normalized to this field value in
the lower panel of Figure \ref{Fig: fraction}.  This plot demonstrates
that there is a very clear transition occurring in the cluster
galaxies between the highest and middle redshift bins, beyond the
global evolution underway in the field.  Indeed, the star-forming
fraction increases monotonically {\it from} the field level, rising
into the cluster cores.

\begin{figure}[bthp]
\epsscale{1.15} % turn on for emulateapj
\plotone{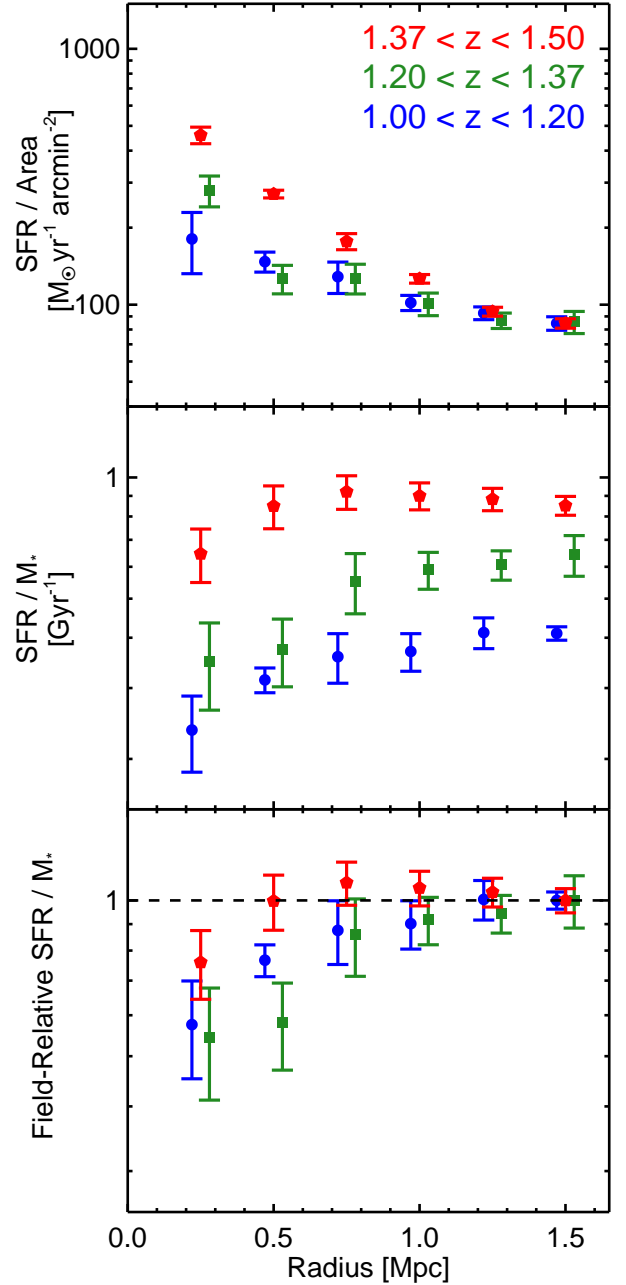}
\caption{SFR surface density ({\it upper panel}) and sSFR ({\it middle
    panel}) vs.~projected clustercentric radius for members with $\log
  (M_\star / \msunmath) \ge 10.1$. The errors are from bootstrap
  resampling.  {\it Lower panel:} Field-relative sSFR, where the field
  values are taken to be the sSFRs at a radius of 1.5 Mpc.  The
  horizontal line illustrates a model with no environment-dependent
  quenching.}
\label{Fig: radial}
\end{figure}

To better explore the star formation properties as a function of
environment, in Figure \ref{Fig: radial} ({\it upper panel}) we plot
the SFR surface density versus clustercentric projected radius.  There
is a strong radial trend, with the SFR surface density increasing by a
factor of 2--3 from the outskirts to the centers of clusters at
$1<z<1.37$.  This is expected even in a quenched environment since the
sheer number of galaxies per unit area is increasing towards the
cluster cores more rapidly than the SFR is falling.

The SFR surface density trend with radius is considerably more
dramatic for clusters in the highest redshift bin ($1.37 < z < 1.50$),
where it increases by a factor of $\sim 5$, reaching a SFR surface
density of nearly $\sim 500$ \msun\ yr$^{-1}$arcmin$^{-2}$ within 250
kpc, \ie\ in the cluster cores.  Building on the rising star-forming
fraction discussed above, this measurement highlights the strong
central star formation occurring in clusters at $z\sim 1.5$.  We
confirm the largely qualitative measurements made on individual,
serendipitously-discovered clusters at similar redshifts; $z=1.46$
\citep{hilton10, hayashi11}, $z=1.56$ \citep{fassbender11_sfr},
$z=1.58$ \citep{santos11} and $z=1.62$ (\citealt{tran10, tadaki12}).

The middle panel of Fig.~\ref{Fig: radial} shows the trend of sSFR
vs.~clustercentric radius.  In the redshift bins at $1<z<1.37$, the
sSFR decreases from a level similar to field galaxies in the outskirts
(see, \eg\ Fig~\ref{Fig: mean_ssfr}) to lower levels toward the
center, a drop of a factor of $\sim 2$.  Although the SFR surface
density increases in the cluster cores at these redshifts, the stellar
mass density is increasing faster, leading to falling sSFR values as
in lower redshift clusters, albeit scaled up by a factor of $\sim 100$
from the local Universe \citep[\eg][]{chung11}.

Conversely, the SFR is increasing so rapidly toward the centers of the
clusters in the most distant redshift bin ($1.37 < z < 1.50$) that the
sSFR is approximately flat right into the cluster cores.  This is the
physically crucial transition, as it indicates the redshift at which
the star formation in recently accreted galaxies has not yet started
to be meaningfully quenched.  These central cluster galaxies are
forming stars as rapidly for their mass as their field galaxy
counterparts.

The radial dependence of the sSFRs in all three redshift bins largely
reflects that seen in the star-forming fractions in Fig.~\ref{Fig:
  fraction}.  This indicates that this activity is happening across
the entire cluster galaxy population, not in just a few star
formation-dominated ULIRGs.  Indeed, as shown in the lower panel of
Figure \ref{Fig: mass_lir}, the bulk of the 24 $\mu$m-detected
population are LIRGs typical of this redshift regime.

In the lower panel of Fig.~\ref{Fig: radial} we plot the
field-normalized radial sSFR trend, where the field sSFR in each
redshift bin is again taken to be the measurement at a radius of 1.5
Mpc.  This removes the strong evolution in field galaxy sSFR, and
allows a quantitative test of the hypothesis that clusters in the
highest redshift bin no longer exhibit environmental quenching.  The
horizontal line illustrates the expected cluster sSFR in this
scenario, identical to the field at all radii.  Comparisons of the
goodness of fit of this hypothesis to the field-normalized cluster
sSFR trends produce $\chi^2_\nu$ values of 6.50, 5.71 and 1.15 for
these sample bins, in order of increasing redshift.  Thus, the
no-quenching model is ruled out for the lower redshift bins at 4.6 and
4.2 $\sigma$, with probability-to-exceed (PTE) values of $4.7 \times
10^{-6}$ and $2.84 \times 10^{-5}$, respectively.  Conversely, the
high redshift bin is a satisfactory fit to the no-quenching model, in
agreement at the 0.97 $\sigma$ level, corresponding to a PTE of 0.31.
Despite this statistical consistency, the sSFR appears slightly
depressed in the innermost radius bin suggesting there may still be a
small amount of quenching in the very center even at this redshift.

\subsection{Cluster-to-Cluster Variation}

The errors on which this statistical test relies are derived from
bootstrap resampling the members in each redshift and radius bin and
therefore reflect the scatter due to population variance in each bin.
To ensure that the abrupt transition in the highest redshift bin is
not due to a single discrepant cluster, we removed each of the 6
highest redshift clusters in turn and recomputed the central sSFR from
the remaining 5 clusters. The scatter from this jackknife process is
smaller than the plotted member-weighted bootstrap errors, confirming
that the transition is characteristic of the cluster sample as a
whole.
 
\begin{figure*}[bthp]
\epsscale{1.15} % turn on for emulateapj
\plottwo{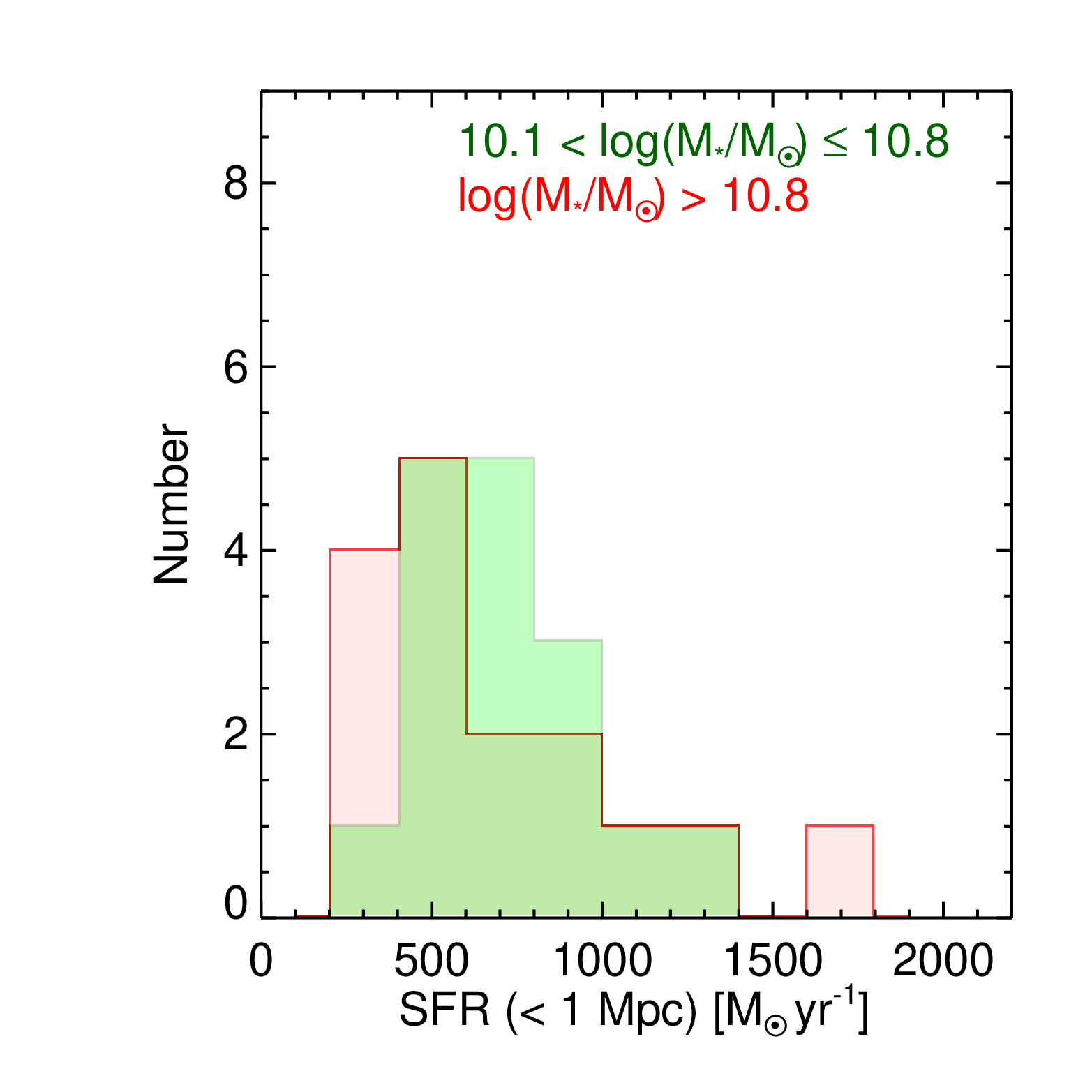}{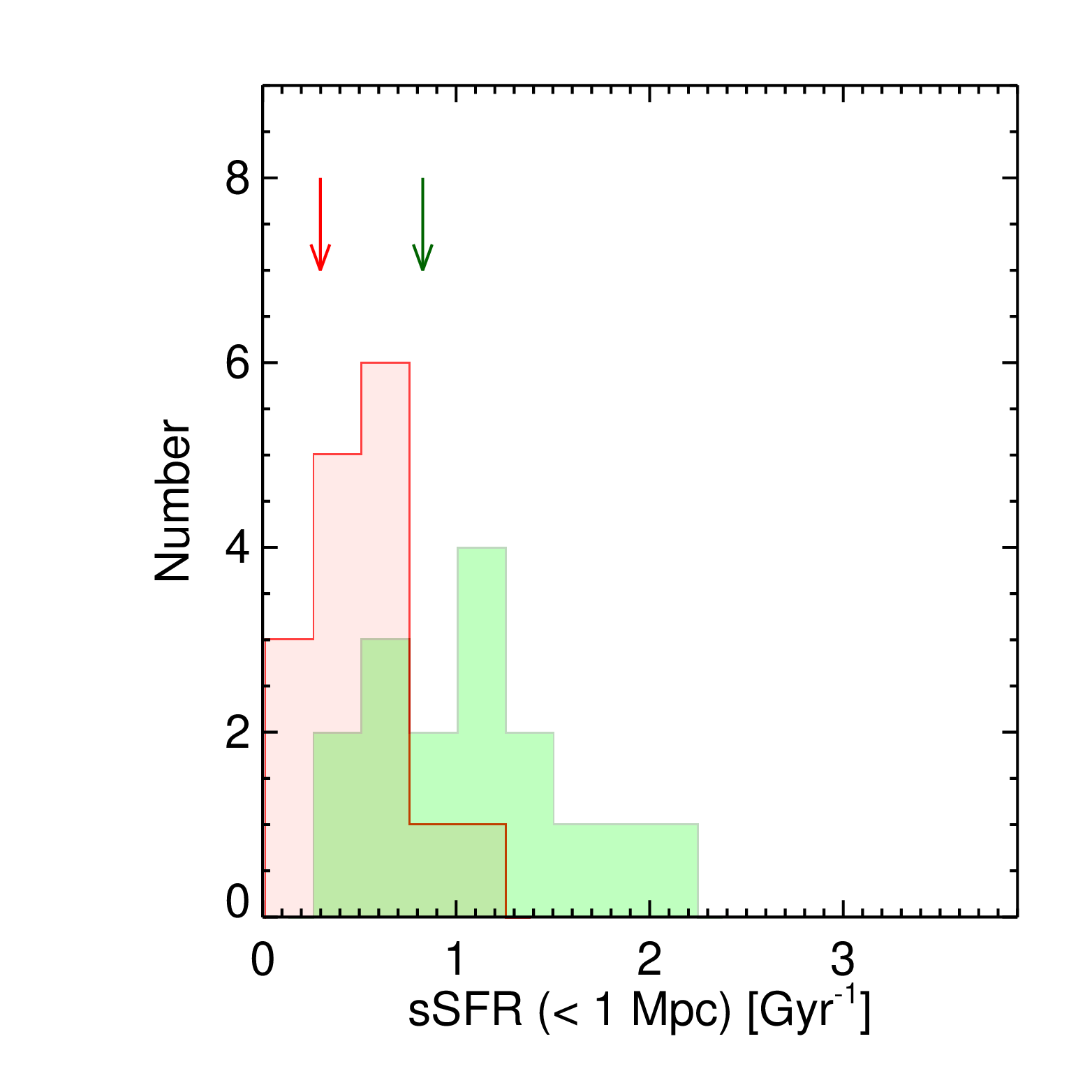}
\caption{Histograms of the total SFR ({\it left}) and sSFR ({\it
    right}) within a projected radius of 1 Mpc, in two bins of stellar
  mass, showing the cluster to cluster variation. The field values of
  the sSFR in the same mass bins, from \citet{santini09}, are
  indicated with arrows.}
\label{Fig: c2c}
\end{figure*}

Figure \ref{Fig: c2c} shows histograms in total cluster SFR ({\it
  left}) and sSFR ({\it right}) in two mass bins for the 16 clusters
in the present sample.  The high-mass and low-mass cluster members
make similar contributions to the total SFR within 1 projected Mpc.
Thus, the enhanced central star formation seen above is occurring in
all galaxies, including the very massive ones.  This is in sharp
contrast to the situation at low redshift, where massive central
galaxies are largely quiescent.  

The sSFR distribution is higher for lower mass galaxies, as seen in
the previous sections. The cluster-to-cluster variation in the SFR and
sSFR, in both mass bins, is a factor of $\sim$3--5.  About half of
this variation is due to the global evolution in the sSFR over the
redshift range probed (Fig.~\ref{Fig: radial}).  For reference, the
field galaxy sSFR values in the same mass and redshift bins, from
\citet{santini09}, are indicated with arrows.

\section{Discussion}
\label{Sec: Discussion}

\subsection{The Era of Star Formation in Galaxy Clusters}
 
These measurements of the star formation properties in high-redshift
ISCS galaxy clusters expand on previous studies in several important
ways.  The sample of 16 IR-selected, spectroscopically confirmed
$1<z<1.5$ clusters is the largest to date.  The SFRs are measured
using mid-IR photometry that is sensitive to the obscured star
formation that dominates the SFR budget, and contaminating AGNs have
been removed using X-ray and mid-infrared methods.  The cluster sample
spans the redshift interval between the era, at $z\la1$, where
clusters cores are less active, and the era at $z \sim 1.5$ where this
trend appears to reverse.  Secure spectroscopic or accurate
photometric redshifts for all members, coupled with multi-wavelength
SED-fitted stellar masses and 24 $\mu$m-based SFRs, allow robust
measurement of the evolution of the SFR and sSFR in high redshift
clusters in bins of redshift, radius and galaxy mass.

We confirm the high fraction of star-forming members in $z\ga 1.4$
clusters reported by previous groups in individual clusters
\citep[\eg][]{tran10, hilton10, lemaux10, hayashi11, fassbender11,
  tadaki12}.  We quantify the star-forming fraction, SFR and sSFR, and
robustly detect a transition between passive, star formation-quenched
low-redshift clusters and relatively unquenched high-redshift clusters
with high central SFRs.  For clusters in the mass range of the ISCS
sample ($\sim 10^{14}$ \msun\ at $z>1$) the transition to the
unquenched star formation era occurs at $z \sim 1.4$.

\subsection{Quenching Mechanisms}
 
The commonly invoked quenching mechanisms in clusters, strangulation
\citep{larson80} and ram pressure stripping \citep{gunn72}, likely
operate at some level in these clusters.  While the quenching
timescale for strangulation ($\sim$several Gyr) is too long to cause
the transition observed in this work, the stripping of the
loosely-bound outer-halo hot gas reservoirs prevents subsequent
fueling and star formation episodes at late times (i.e.~at
$z<1$). 

Ram pressure stripping can remove tightly bound disk gas on relatively
rapid timescales ($\sim 1$ Gyr), particularly in $z>1$ clusters in
which the dynamical time is fairly short.  As such, it can rapidly
quench star formation in cluster galaxies.  However, the ram pressure
goes as the square of the orbital velocity, and hence is more
effective in massive, high-dispersion ($\ga 1000$ km s$^{-1}$)
clusters than in typical $z>1$ ISCS clusters, which have more modest
dispersions ($\sim 700$ km s$^{-1}$; \citealt{brodwin11}).  Further,
detailed simulations of ram pressure stripping suggest that at least
30\% of a galaxy's disk gas remains unstripped 10 Gyr after initial
infall \citep{mccarthy08}.  Therefore, while this mechanism may be
responsible for a portion of the quenching, it likely cannot fully
explain the strong quenching occurring over $z=1.5 \rightarrow 1.0$
(Figs.~\ref{Fig: fraction} and \ref{Fig: radial}). In the more passive
era following the one studied in the present work, over $z\sim 1.0
\rightarrow 0.3$, \citet{alberts13} observe a gradual, continuous
quenching of star formation in ISCS clusters.  They suggest this is
likely a due to a combination of strangulation and ram pressure
stripping.

The transition at $z\sim 1.4$ is strikingly similar to the recent
results of \citet[][see their Fig.~7]{mancone10}.  That work measured
rest-frame infrared luminosity functions for the full ISCS cluster
sample, consisting of 335 clusters over $0.3<z<2$.  Using the same
accurate photometric redshifts as in the present work, the evolution
in the cluster luminosity function was mapped out at both 3.6\mum\ and
4.5\mum.  At $z < 1.3$ the evolution in the characteristic magnitude
$M^*$, an extremely good proxy for stellar mass given the rest-frame
NIR sampling, was fully consistent with the passive evolution model
found in most other studies \citep[\eg][]{stanford98}.  However, at $z
\ga 1.4$, \citet{mancone10} found an abrupt $\sim 1$ mag dimming of
$M^*$ in the cluster luminosity functions, corresponding to a stellar
mass growth of a factor of $\sim 2-4$ from $z\sim 2$ to $z \sim 1.3$.
This was interpreted as evidence of mass assembly via merging in these
high-redshift cluster galaxies.  

Evidence that mergers may play an important role in the evolution of
galaxy populations in distant clusters has been accumulating.
Luminosity functions presented by several groups exhibit a paucity of
massive ($ > L^*$) member galaxies on the red sequence at $z \ga 1.4$
\citep{hilton09, hilton10, fassbender11, rudnick12, mancone12}.
Direct and indirect evidence for a sharply increased merging rate, a
factor of 3--10 higher than in contemporaneous field galaxies, has
been seen in a $z=1.62$ cluster \citep{lotz13,rudnick12}.  Evidence
for a stochastic star formation history, with young early-type
galaxies (presumably formed via mergers) continuously arriving on the
cluster red sequence at $1<z<1.5$, has been reported by
\citealt{snyder12} (see also \citealt{jaffe11}). A rapid
two-order-of-magnitude increase in the fraction of AGN in clusters at
$z\sim 1.5$ is reported by \citet{martini13}.  Finally, a high
fraction of post-starbust central galaxies are detected in somewhat
lower redshift ($0.6 \la z \la 1$) clusters
\citep{poggianti09,muzzin12}.

From the \spitzer/IRAC data at 3.6\mum\ and 4.5\mum, \citet{mancone10}
could not discern whether this epoch of assembly in ISCS clusters
consisted of mergers that were ``wet'' (\ie\ collisional mergers of
gas-rich galaxies, triggering a starburst and fueling black hole
accretion via nuclear inflow of gas; \eg\ \citealt{barnes91};
\citealt{springel05}; \citealt{hopkins06}; \citealt{narayanan10}) or
``dry'' (collisionless mergers of gas-free galaxies; \eg\
\citealt{vandokkum05}; \citealt{bell06}).  With the longer wavelength
MIPS data, we now have at least circumstantial evidence that a
substantial fraction of the mergers inferred by that work are likely
inducing massive starbursts.  Visual inspection of several of the
highest \lir\ galaxies in high-resolution \hst\ images shows a large
number of disturbed and/or merging systems.  This evidence for ``wet''
mergers corroborates the findings of \citet{desai11} that low redshift
elliptical galaxies have residual 24 $\mu$m emission, suggestive of
past collisional mergers.  A complete statistical description of the
star formation properties of a morphological, merger-selected sample
of ISCS cluster members will be presented in a future paper.

If these observed starbursts are merger-induced, recent simulations
\citep[\eg][]{springel05, hopkins06,narayanan10} predict that AGN
feedback may also play a significant role in quenching the star
formation in these cluster galaxies.  In these models, a merger of
gas-rich progenitors triggers both massive starbursts and fuels a
powerful central AGN.  The AGN heats and expels the remaining gas,
leading to a rapid quenching of star formation, on $\sim 100$ Myr
timescales.  This model helps explain the transition observed in the
ISCS cluster galaxies and provides a mechanism that allows them to
appear to be passively evolving only $\sim 1$ Gyr later.  The most
massive merger-induced starbursts will likely also experience
significant feedback from supernovae and strong stellar winds which
can efficiently expel gas, particularly in the outer regions of the
galaxies \citep{diamond-stanic12}.

The observed starbursts are not likely to be driven by cold-mode
accretion \citep[\eg][]{keres05,dekel06,dekel09,nelson13} as these
cluster galaxies reside within a hot ICM that should prevent cold
streams to all galaxies except possibly the BCG.  Further, this
scenario offers no straightforward way to rapidly quench the star
formation for sub-L$^*$ galaxies.  The cold-stream only shuts off when
the halo reaches a mass large enough ($M_{\rm \small halo} \ga
10^{12}$ \msun) to shock heat the infalling gas.  Indeed, the
simulations of \citet{keres05} show that cold flows are only important
in areas of low galaxy density.  In groups or clusters the
contribution of cold-mode accretion is expected to be negligible.

\subsection{A Model for Galaxy Cluster Evolution}

The standard cluster formation paradigm explains many of the observed
properties of cluster galaxies.  It holds that both the primordial
cluster seed galaxies and those accreted from the field are stripped
of their hot, loosely-bound gaseous halos by the ICM.  Over a
dynamical time ($\la 1$ Gyr in high-redshift clusters) ram pressure
stripping removes of order half of the cool gas from the galaxy disks
\citep{mccarthy08}.  Cold mode accretion is inefficient in hot cluster
halos \citep{keres05}, so in the absence of mergers secular star
formation ceases when the remaining cool gas supply in the galaxy
disks is exhausted.  At this point the galaxies become quiescent and
evolve passively thereafter, becoming ``red and dead'' by the present
day.  While many elements of this model are probably correct, it does
not explain the extensive star formation (this work;
\citealt{snyder12, zeimann13, alberts13}), merger \citep{mancone10}
and AGN activity \citep{galametz10, wagg12, martini13} observed to be
taking place in the ISCS clusters at $1<z<1.5$.

We find evidence for an additional, significant epoch of merging
activity taking place in clusters at $z \ga 1.4$, which is also the
era of peak star formation and AGN activity in the general galaxy
population.  This merging epoch, observed statistically by
\citet{mancone10} in the rest-frame near-IR cluster luminosity
functions, is occurring between gas-rich progenitors and leads to
vigorous starbursts that we detect in the mid-IR.  The resulting SFRs
in some galaxies are so high they would, if allowed to proceed
unquenched, consume the bulk of the cold gas remaining in these
cluster galaxies on a very short timescale ($\sim 100-300$ Myr).  The
mergers also feed the accretion of central black holes.  When these
black holes enter an active AGN phase, they heat and/or expel the
remaining cold gas, abruptly quenching the star formation.

This model explains the bulk of the observations of cluster evolution
to date.  In particular, it offers a more physically motivated
explanation for the apparent pure passive evolution seen in $z<1$
cluster studies \citep[\eg][]{stanford98,vandokkum07, muzzin08}.
These studies typically employ models in which the last significant SF
activity occurred at $z>2$. These models are ruled out by recent
observations of vigorous star formation in high redshift clusters,
most dramatically in the present paper.  While a passive model fit the
\citet{mancone10} observations for clusters at $z \la 1.3$, it failed
completely at higher redshift, where the galaxies were substantially
less massive than expected in a merger-free passive model. Similarly,
the rapid reddening observed in E08 at $z \sim 1.4$ (their Fig.~19) is
better explained by an epoch of merger-driven obscured star formation
than by a sudden change in the passive-model formation redshift from
$z_f \sim 4$ to $z_f \sim 30$. In the AGN-quenched model, cluster
galaxies will have faded, reddened and appear ``red and dead'' by $z
\sim 1$ (1-2 Gyr after quenching).  If the galaxies evolve passively
thereafter, they will appear in the present day to have
luminosity-weighted pure passive-model formation redshifts of $2 \la z
\la 4$.  That is an average of the more extended and punctuated star
formation history, from formation at $z\ga 4$ to final starburst
ending at $z \sim 1.4$.

Indeed, recent analyses of the colors and scatters of red-sequences in
high-redshift clusters, such as \citet{jaffe11} and \citet{snyder12},
have tested models with ongoing stochastic or even continuous star
formation, ending $1-2$ Gyr prior to the epoch of observation.  They
find good fits for models in which the interval between formation and
final placement on the cluster red-sequence is similar to the
timescale for AGN-quenching in mergers.  These delayed models, in
which the last big epoch of star formation occurs at $z<2$ but is
complete by $z \sim 1.2$, are qualitatively consistent with the ISCS
cluster observations and the picture we have presented to describe
them.

Unlike explanations of low-redshift galaxy properties, such as the
black hole-bulge mass relation \citep[\eg][]{ferrarese00} or the very
red colors of the most massive field galaxies \citep[\eg][]{croton06},
mergers in this work were not merely invoked as a useful mechanism to
explain the observations.  The era of significant merging was first
{\it observed} (statistically) in \citet{mancone10}, and in the
present work we directly observe in the same clusters the vigorous
starbursts expected from gas-rich mergers.  Further, we indirectly
observe the rapid quenching of that star formation expected due to
feedback from the central AGN.  The AGN feedback scenario offers a
natural explanation for all these observations and, furthermore, makes
several falsifiable predictions.

Most directly it predicts a strong increase in the incidence of AGN
activity in clusters at $z>1$ compared with those at lower redshifts.
Evidence of this has already been seen by several groups
\citep[\eg][]{martini09,kocevski09a,lemaux10,fassbender12,tanaka13}
including our own statistical analysis of the full ISCS sample
\citep{galametz09} in which we find that X-ray-selected AGN are at
least 3 times more prevalent in clusters at $1<z<1.5$ compared with
clusters at $0.5<z<1$.  Since cold flows are inefficient in hot
cluster halos, the role of mergers in rich environments is likely to
be even more important than in the field.  We therefore might expect
not only a rapid increase in the incidence of AGN activity in clusters
with increasing redshift, but an increase that is significantly {\it
  more} rapid than is occurring in the field.  A detailed new analysis
of the clusters in this work \citep{martini13}, using deep X-ray data
and extensive spectroscopy, confirms this is the case.  Although the
AGN fraction is $\approx 6$ times higher in the field than in clusters
in the local Universe, the fractions are comparable at $z \sim 1.25$.
\citet{martini13} conclude that this differential evolution of the AGN
fraction in the field and clusters is strong evidence for
environment-dependent AGN evolution.

This model also corroborates the findings of \citet{brodwin08} that
the brightest Dust-Obscured Galaxies \cite[DOGS,][]{dey08,pope08},
which are dusty AGN-dominated ULIRGs at $1.5 \la z \la 2.5$ (also see
\citealt{farrah06, magliocchetti08, starikova12, viero13}), have
similar clustering properties to galaxy groups and are located in
rare, rich environments.  Although this extreme population is excluded
from the present study by our AGN rejection and the limited redshift
overlap with the clusters in this sample, this evolutionary
relationship between DOGs and clusters is an interesting and important
topic that will be addressed in a future paper (Williams \etal\ in
prep.).
 
In addition to rendering cluster galaxies largely quiescent at $z \la
1$, in this model cluster galaxies at such redshifts should show signs
of both recent starburst activity and of rapid AGN-driven quenching.
This is seen in a several studies, with high post-starburst and
low-level AGN fractions in clusters at $0.8 \la z \la 1$
\citep[\eg][]{kocevski09b,lemaux10, muzzin12}.

\subsection{Epoch of Merging in ISCS Clusters}

Mergers are most efficient when galaxy space densities are high and
relative velocities are low.  In the local Universe, group
environments, with their relatively high source densities and modest
velocity dispersions, are expected to have the highest merging
frequency \citep{hopkins08}.  Dispersions are too high ($> 1000$ km
s$^{-1}$) in present-day massive clusters ($M \sim 10^{15}$ \msun) to
produce much merging.  But the $z >1$ progenitors of these massive
clusters had smaller halo masses and velocity dispersions, and higher
densities of galaxies with extended gas-rich disks, all of which led
to a higher merging efficiency.  For clusters with masses typical of
the ISCS sample ($\sim 10^{14}$ \msun\ at $z>1$), major merging had
likely been occurring continuously since initial formation, but should
have begun to subside by $z\sim 1.5$ due to ever-growing velocity
dispersions. Indeed, in their long-baseline \herschel\ stacking study,
\citet{alberts13} only find evidence for substantial merging in ISCS
clusters at $z \ga 1.4$.  An enhanced merger rate is also directly
observed in ClG~J0218.3-0510 at $z=1.62$ \citep{lotz13} and inferred
by \citet{rudnick12} from the evolution of its luminosity function.

If this model is correct, more massive high-redshift clusters such as
SPT-CL J0205-5829 at $z = 1.32$ \citep{stalder13}, XMMU J2235.3-2557
at $z=1.39$ \citep{mullis05,rosati09} and SPT-CL J2040-4451 at $z =
1.48$ \citep{bayliss13}, with masses of $\sim 9 \times 10^{14}$~\msun,
$\sim 6 \times 10^{14}$~\msun\ and $\sim 6 \times 10^{14}$~\msun,
respectively, should no longer be experiencing efficient merging due
to their high in situ velocity dispersions.  Rather their transition
redshifts, when phase space conditions were more conducive to major
merging activity, should be considerably higher than that seen in the
ISCS. Indeed, these clusters have relatively low central star
formation rates \citep{stalder13,grutzbauch12,bayliss13,santos13},
$\la 5\times$ below those in the present work, consistent with already
being largely quenched and passive in their cores.  A related
prediction is that the scatter in the colors of red-sequence galaxies
in these very massive clusters should be smaller than that measured by
\citet{snyder12} for ISCS clusters at similar redshifts.

Another test of this prediction is forthcoming, using the massive
cluster \cl\ at $z=1.75$ \citep{brodwin12, gonzalez12, stanford12}.
This cluster is, in an evolutionary sense, a precursor of these three
massive clusters and of the most massive clusters at all redshifts,
including Coma.  Its star formation properties, measured from deep
\herschel\ observations, will be presented in an upcoming paper
(Alberts et al.~in prep).  Though very massive for its redshift
($\Mtwo \sim 4 \times 10^{14}$ \msun) it is very compact, with the
majority of the infrared overdensity within a projected radius of
$\sim 30\arcsec$.  As such, conditions may be still be suitable for
substantial merging and merging-induced starbursts.

\section{Conclusions}
\label{Sec: Conclusions}

We have investigated the star formation properties of 16
high-redshift, IR-selected galaxy clusters from the ISCS.  Using deep
\spitzer\ 24 \mum\ imaging, we characterized the obscured star
formation in these clusters as a function of redshift, stellar mass
and clustercentric radius.  For 6 of these clusters, including 5 at
$z>1.35$, we also provide the ground-based spectroscopic confirmation.
Redshifts from the \hst/WFC3 grism, along with a complementary
analysis of the unobscured H$\alpha$ star formation activity, are
presented in \citet{zeimann13}.

The primary result is that $z > 1$ ISCS clusters have substantial star
formation activity occurring at all radii, including in the cluster
cores.  The SFRs in these cluster galaxies are similar to that of
field galaxies at similar redshifts, suggesting that we are probing
the era before cluster quenching was complete.  As we have
conservatively rejected X-ray and IR AGN from this study, these
cluster star formation rates are lower limits.  

The transition between the low redshift ($z<1$) era, in which cluster
galaxies are significantly quenched relative to the field, and the era
of cluster formation, in which cluster galaxies form stars at the same
rate as field galaxies for their masses, occurs at $z\sim 1.4$ in the
ISCS sample.  Below this redshift, although significant star formation
occurs in clusters at all radii, the sSFR drops near the core,
suggesting active environmental-dependent quenching.  At redshifts
above $z\sim 1.4$, there is evidence from both the fraction of
star-forming galaxies and the sSFR that quenching in the cores is
minimal.  Above this redshift, cluster galaxies are forming stars at
the rate expected for field galaxies of similar mass, independent of
their location in the cluster.

There is a factor of $\sim$3--5 variation in the star formation
activity from cluster to cluster in this IR-selected sample.  About
half of that variation is due to the observed redshift evolution, but
the rest is intrinsic scatter in the population.  This variance
suggests that substantially larger samples will be required to improve
upon the present work.  In particular, single-cluster studies are
difficult to interpret and to place in a meaningful evolutionary
context.

Combining the present measurements with recent independent results
from the ISCS survey, such as the strong increase in AGN density
\citep{martini13}, the stochastic star formation histories
\citep{snyder12, alberts13}, and the statistical evidence for a
significant assembly epoch at $\sim 1.4$ \citep{mancone10}, we suggest
that mergers likely play a significant role in the observed star
formation activity.  In addition to plausibly inducing the large
starbursts seen in these cluster galaxies, merger-fueled AGN feedback
\citep[\eg][]{hopkins06} may naturally explain the rapid truncation of
star formation that occurs between $z\sim 1.5$ and $z \la 1$, by which
time the cores of clusters become largely quiescent
\citep[\eg][]{vulcani10,finn10,muzzin12} with high post-starburst
fractions \citep{poggianti09, muzzin12}.

If mergers do play a significant role in the transition between the
unquenched and quenched eras, the redshift of this transition is
likely strongly dependent on cluster halo mass.  Mergers require
relatively low velocity dispersions, so a prediction of this work is
that more massive clusters than those in the ISCS sample would
experience this transition at higher redshifts.  Studies of the star
formation properties in the few known high-mass, high-redshift
clusters \citep{stalder13, grutzbauch12, bayliss13,santos13} support
this expectation.

\acknowledgments This work is based in part on observations made with
the {\it Spitzer Space Telescope}, which is operated by the Jet
Propulsion Laboratory, California Institute of Technology under a
contract with NASA. Support for this work was provided by NASA through
an award issued by JPL/Caltech.  Support for HST programs 10496,
11002, 11597, and 11663 were provided by NASA through a grant from the
Space Telescope Science Institute, which is operated by the
Association of Universities for Research in Astronomy, Inc., under
NASA contract NAS 5-26555.  This work is based in part on observations
obtained with the {\it Chandra X-ray Observatory} ({\it CXO}), under
contract SV4-74018, A31 with the Smithsonian Astrophysical Observatory
which operates the {\it CXO} for NASA.  Support for this research was
provided by NASA grant G09-0150A.  This work is based in part on data
obtained at the W.~M.~Keck Observatory, which is operated as a
scientific partnership among the California Institute of Technology,
the University of California and the National Aeronautics and Space
Administration.  The Observatory was made possible by the generous
financial support of the W.~M.~Keck Foundation.  This work makes use
of image data from the NOAO Deep Wide--Field Survey (NDWFS) as
distributed by the NOAO Science Archive. NOAO is operated by the
Association of Universities for Research in Astronomy (AURA), Inc.,
under a cooperative agreement with the National Science Foundation.

We are grateful to the referee for a helpful report that improved the
clarity of the paper.  We thank P.~Santini for providing her
data in a digital form.  We appreciate several useful conversations
with C.~Papovich, M.~Cooper, M.~Dickinson, N.~Reddy and S.~Salim.
This paper would not have been possible without the efforts of the
support staffs of the \spitzer\ Space Telescope, {\it Hubble} Space
Telescope, \chandra\ X-ray Observatory and W.~M.~Keck Observatory.

\bibliographystyle{astron2} \bibliography{bibfile}

\end{document}